\RequirePackage{fix-cm}
\documentclass[twocolumn,final=true]{svjour3}          
\smartqed  
\usepackage{graphicx}

\usepackage{mathptmx}      

\title{The Unit-B Method --- Refinement Guided by Progress Concerns
   \thanks{This is an extended
    version of \cite{ifm/Hudon/Hoang13}}
}
\titlerunning{The Unit-B Method}

\author{Simon Hudon \and Thai Son Hoang \and Jonathan S. Ostroff}
\authorrunning{S.~Hudon \and T.S.~Hoang \and J.S.~Ostroff}
\institute{
  Simon Hudon \at
  Electrical Engineering \& Computer Science,  York University,
  Toronto, Canada \\
  \email{simon@cse.yorku.ca} \and
  Thai Son Hoang \at Institute of Information Security, ETH Zurich, Switzerland \\
  \email{htson@inf.ethz.ch} \and
  Jonathan S. Ostroff \at Electrical Engineering \& Computer Science,
  York University, Toronto, Canada \\
  \email{jonathan@cse.yorku.ca}.
}

\date{\today}

\newcounter{def}
\newtheorem{Definition}[def]{Definition}

\newcounter{ptl}
\newtheorem{Postulate}[ptl]{Postulate}
\newcommand{\Post}{Postulate}

\newcounter{thm}
\newtheorem{Theorem}[thm]{Theorem}
\newcommand{\Thm}{Theorem}
\newcounter{cor}
\newtheorem{Corollary}[cor]{Corollary}
\newcommand{\Cor}{Corollary}

\usepackage{xspace}
\usepackage{color}
\usepackage{datetime}

\usepackage{amsmath}
\usepackage{amssymb}

\usepackage{space}

\usepackage{varioref}

\usepackage{times}

\usepackage{wrapfig}

\usepackage{calculation}

\usepackage{comp-cal}

\usepackage{rel-cal}

\usepackage{seqcal}

\usepackage{xspace}

\usepackage{ulem}

\newcommand{\msout}[1]{\text{\sout{\ensuremath{#1}}}}

\usepackage{unity}

\usepackage{bsymb}

\usepackage[color, nobox]{eventB}

\usepackage{sgnctrl}

\usepackage{tla}

\usepackage{unitb}

\usepackage{tikz}
\usetikzlibrary{snakes,arrows}

\usepackage[compact]{reqdoc}

\usepackage{minibox}

\usepackage[plainpages=false]{hyperref}
\hypersetup{
  pdftitle={},
  pdfauthor={Simon Hudon, Thai Son Hoang, and Jonathan S. Ostroff
    Department of Computer Science, York University, Toronto, Canada
    and
    Institute of Information Security, ETH Zurich, Switzerland
  },
  pdfsubject={Refinement, Progress},
  pdfkeywords={Refinement, Progress, Temporal Logic},
  colorlinks=true
}

\begin{document}

\maketitle


\begin{abstract}
  We present Unit-B, a formal method inspired by Event-B and UNITY.
  Unit-B aims at the stepwise design of software systems satisfying
  safety and liveness properties. The method features the novel notion of coarse and fine schedules, a
  generalisation of weak and strong fairness for specifying events'
  scheduling assumptions.  Based on events schedules, we propose proof rules to reason about progress properties and  a refinement order  preserving both liveness and safety
  properties.  
  We illustrate our approach by an example to show that
  systems development can be driven by not only safety but also liveness
  requirements.

\end{abstract}

\section{Introduction}
\label{sec:introduction}

Developing systems satisfying their desirable properties is a non-trivial
task.  Formal methods have been seen as a possible solution to the
problem.  Given the increasing complexity of systems, many formal
methods adopt refinement techniques, where systems are developed
step-by-step in a property preserving manner.  In this way, a system's
details are gradually introduced into its design within a hierarchical
development.

System properties are often categorised into two classes:
\emph{safety} and \emph{liveness}~\cite{DBLP:journals/tse/Lamport77}.
A safety property ensures that undesirable behaviours will never
happen during system executions.  A liveness property guarantees that
eventually desirable behaviours will happen.  Ideally, systems should
be developed in such a way that they satisfy both their safety and
liveness requirements.  Although safety properties are often
considered the more important ones, we argue that having \emph{live}
systems is also important.  A system that is safe but not live can be
useless.  For example, consider an elevator system that does not move.
Such an elevator system is safe (nobody gets hurt), yet worthless.
According to a survey~\cite{DBLP:conf/icse/DwyerAC99}, liveness
properties (in terms of \emph{existence} and \emph{progress}) amount
to 45\% of the overall system properties.

\subsection{Motivation} 
In many refinement-based development methods (e.g.,
B~\cite{abrial96:_b}, Event-B~\cite{DBLP:books/daglib/0024570},
VDM~\cite{DBLP:books/daglib/0068091},
Z~\cite{DBLP:books/daglib/0068766}), the focus is on preserving safety
properties.  A common problem for such safety-oriented methods is that
when applying them to the design of a system, it is possible to make
the design so safe that it becomes unusable.  This would happen if we
strengthened the guards of the events (in \eventB) or choose strong
preconditions (in B, Z, VDM) to facilitate the proof of safety
properties but in such a way that, in cases where the operations or
events are needed to make the system progress, they are not enabled.
Concretely, in an elevator system, this might result in a controller
which eventually stops opening the door to the  elevator (possibly
despite there being people inside) in order to satisfy the safety
property that the door not be opened between floors. 
It is hence our aim to design a refinement framework preserving both
safety and liveness properties.

\unity ~\cite{DBLP:books/daglib/0067338} has a  calculus for liveness 
but does not support refinement of programs.  Specifications are 
written in the \unity logic (a subset of temporal logic) and 
implementations are programs (or transition systems). The initial 
specification can be refined by a stronger set of temporal properties 
but once the temporal properties are implemented, further refinement 
of the programs is not possible. 

Event-B \cite{DBLP:books/daglib/0024570} has a calculus for refinement
of safety properties, but does not provide much support for liveness
and fairness. Instead, Event-B provides the notion of convergence. In
a system, a set of events are convergent if they cannot prevent the
other events from happening.  This can be used, for instance, to
develop model of a sequential program, to prove that the program
terminates.  Convergence is proven by choosing a variant for the
system, i.e., an expression whose type is well ordered (e.g., natural
numbers or finite sets).  Then, it must be proved that all convergent
events are guaranteed to decrease the variant whenever they are
executed.

In Event-B, liveness properties cannot be directly expressed and
proved. One justifies the validity of a liveness property (e.g.
$\square \lozenge \Bevt{evt}$, i.e., infinitely often, event $\Bevt{evt}$
occurs) by showing that the system is deadlock free and that events
other than $\Bevt{evt}$ are convergent. However, one must show that deadlock
freedom is preserved in each following refinement, and that all new
events are convergent.  If a spontaneous event (i.e. non-convergent)
is needed in a refinement (e.g., an event representing an environmental
action), liveness is no longer preserved.  Also, only one liveness
property per system can be supported.

Our \unitb method~\cite{thesis/hudon2011} is inspired by the treatment of
liveness in \unity and refinement in \eventB.  It improves on both
methods by offering a notion of refinement that preserves liveness
applicable to reactive and distributed systems.  It does this by the
introduction of coarse and fine schedules on events and event indices.

\hyphenation{pro-gress}
In the subsequent, we present a small example to contrast \eventB's safety-based
style of reasoning with \unitb's liveness-driven style. We present two
high level models (one in \eventB and the other in \unitb) of a mutual exclusion protocol. In each model, we show the important
safety and liveness properties that one can prove. More specifically,
we study three requirements (1) mutual exclusion (safety), (2) minimal
progress and (3) individual progress.

\subsubsection{An Event-B Model}
\newBvrb{st}
\newBcst{waiting}
\newBcst{idle}
\newBcst{cs}
\newBset{Pcs}
\newBevt{request}
\newBevt{enter}
\newBevt{exit}
\begin{figure}
\begin{align*} 
  &\variables{\st } \\
  &\invariants{
     ~\Binv{inv0}:& \st \in \Pcs \tfun \{\idle,\waiting,\cs\}
  \\ ~\Binv{inv1}:& 
        \qforall{p,q}{p \neq q}
          { \neg(\st.p = \cs \land \st.q = \cs) } } \\
  &\Bevents:\\
  & \quad\request~\widehat{=}~\Bstatus~ convergent \\
  & \quad \quad \Bany ~p~ \Bwhere ~\st.p = \idle~ \Bthen
  ~\st.p \bcmeq \waiting~ \Bend \\[1ex]
  & \quad \enter~\widehat{=} ~\Bstatus~ordinary \\
  & \quad \quad \Bany ~p~ \Bwhere ~\st.p = \waiting \land
  \qforall{q}{p \neq q}{\st.q \neq {\cs}}~ \Bthen\\
  & \quad \quad \quad \st.p \bcmeq {\cs}\\
  & \quad \quad \Bend \\[1ex]
  & \quad \exit~\widehat{=} ~\Bstatus~ convergent \\
  & \quad \quad \Bany ~p~ \Bwhere ~\st.p = \cs~ \Bthen
  ~\st.p \bcmeq \idle~ \Bend
\end{align*}
Note: $p,q \in \Pcs$ is implicitly assumed.
\caption{Event-B mutual exclusion specification}
\label{fig:event-B:mutex}
\end{figure}

The \eventB model, in Figure~\ref{fig:event-B:mutex}, formalises a set
of processes ($\Pcs$) each of which is in one of three states:
\idle, \waiting, and \cs (i.e., in their \emph{critical section}). The state of every process is recorded in the (global) variable
$\st$ (see invariant $\Binv{inv0}$). The safety requirement of the protocol, that
of mutual exclusion, is captured by invariant $\Binv{inv1}$ and can be proved
at this level of abstraction.

In order to reason about liveness in this \eventB system, we need a
variant. The variant is chosen on the basis of the exact property that
we want to demonstrate. We are interested in proving continuous
progress, i.e., as long as there are processes waiting to enter their
critical section, some process will get to enter. In linear time
temporal logic, this is formulated as:
\begin{align}
\tag{\Binv{prg0}}
\square (~\qexists{p}{}{\st.p = \waiting} 
  \1\implies \lozenge \qexists{p}{}{\st.p = \cs}~) \label{intro:livelock}
\end{align}
In this property, the $p$ in $\st.p = \waiting$ and the $p$ in $\st.p =
\cs$ are not necessarily the same and  individual processes might wait
forever. There are two ways in which an execution of the \eventB
model might fail to satisfy this (weak) liveness property:

\begin{enumerate}
\item The system executes forever but, after a point where some processes 
  are waiting, only events $\request$ and $\exit$ are taken.
\item The system deadlocks, i.e. terminates, in a state where some 
  processes are still waiting.
\end{enumerate}

Issue 1 can be addressed by using the following variant and by making
events $\request$ and $\exit$ convergent (as in Figure~\ref{fig:event-B:mutex}).
\[  \variant{ 
      2 \times \qcount{p}{}{\st.p = \cs} \2+ 
      \qcount{p}{}{\st.p = \idle} } \] 
In the above expression, the notation $\qcount{x}{R}{T}$, the counting
quantifier, is used to designate the number of values of $x$ that
satisfy $T$ given that they satisfy $R$. Convergent events are
required to decrease this natural number variant.
Event $\exit$ decreases the first term by 2 and increases the second
term by 1 while event $\request$ decreases the second term and leaves
the first term unchanged. The two events are therefore convergent.

This line of reasoning proves that any (possibly partial) execution of
the system where $\enter$ is not executed must be finite.
It also follows that any infinite execution of the model includes an
infinite number of occurrences of $\enter$. In this case,
it means that $\enter$ has to occur infinitely many times in infinite
executions.

Issue 2 is addressed by ensuring that, at any time, there is at least
one enabled event. This is known as a proof of \emph{deadlock freedom}.
\begin{align*} 
\begin{array}{rll} 
\Bthm{dlf}:\qquad
 		& \qexists{p}{}{\st.p = {\idle}} & \text{// request}
\\ \lor~& \qexists{p}{}{
	\begin{array}{rl}
	   & \st.p = {\waiting}
	\\ \land & \qforall{q}{q\neq p}{\st.q \neq {\cs}}
	\end{array} 
	}& \text{// enter}
\\ \lor~& \qexists{p}{}{\st.p = {\cs}} & \text{// exit}
\end{array}
\end{align*}
%

In this small system, deadlock freedom is easy to prove. However, the
size of its formulation grows with the number of events of the system
and it cannot, in general, be broken down into smaller proof
obligations.


In addition, the (weak) liveness property \eqref{intro:livelock} is
not automatically satisfied by refinements of the system. In order to
preserve it, we need to make convergent all the events introduced in
successive refinements and we need to prove relative deadlock freedom
at each level of abstraction, a burden that only grows more daunting
as a development progresses.

As mentioned earlier, this does not prove individual progress of the
processes involved in the protocol. We would like to prove the
following (strong) liveness property:
\[ \qforall{p}{}{ \square(~ \st.p = \waiting \1\implies \lozenge \st.p = \cs ~) }
\]
i.e., every process waiting eventually enters its critical section. It
is not possible to prove such individual progress using the \eventB
model of Figure~\ref{fig:event-B:mutex}. In order to do such a
proof, we would need to include in the model a description of a
scheduler. In other words, a low level design is necessary even for a
high level liveness property. This is contrary to the idea of
refinement: properties should be provable at the level of abstraction
and the  level of details to which they pertain. This is what \unitb
accomplishes.


\paragraph{Notational Convention.}

The examples in this paper rely heavily on discrete mathematics and
predicate calculus. With the exception of function application, we
borrow the set-theoretic and relational notation from the \eventB
book~\cite{DBLP:books/daglib/0024570}; function application, written
$f.\,x$ with $f$ the function and $x$ the argument, as well predicate
calculus and generalized quantifier notation are taken from E.W.
Dijkstra \cite{EWD:EWD1300}. 

In Dijkstra's relativised quantifier notation, $\qforall{x}{R}{T}$ and
$\qexists{x}{R}{T}$, with $R$ the range of the quantifications and $T$
the term, are equivalent to the  more common $(\forall x ~\bullet~ R
\implies T)$ and $(\exists x ~\bullet~ R \land T)$. 
%
%
The notation for Temporal Logic is taken from
\cite{DBLP_books_aw_Lamport2002}. 

\subsubsection{A \unitb Model}
\label{sec:unitb-model}
Figure~\ref{fig:unit-b:mutex} shows a \unitb model for the same
problem as Figure~\ref{fig:event-B:mutex}. The \unitb has the same set
of variables and invariants as the \eventB model from
Figure~\ref{fig:event-B:mutex}. Figure~\ref{fig:unit-b:mutex} only
shows the events of the \unitb model.

\begin{figure}
  \begin{align*}
    & \Bevents:\\
    & \quad\request~[p]~\widehat{=}~ \\
    & \quad \quad \Bwhen ~\st.p = \idle~ \Bthen
    ~\st.p \bcmeq \waiting~ \Bend \\[1ex]
    & \quad \enter~[p]~\widehat{=} \\
    & \quad \quad \Bduring ~\st.p = \waiting~ \Bupon
    ~\qforall{q}{p \neq q}{\st.q \neq {\cs}}~ \Bbegin\\
    & \quad \quad \quad \st.p \bcmeq {\cs}\\
    & \quad \quad \Bend \\[1ex]
    & \quad \exit~[p]~\widehat{=} \\
    & \quad \quad \Bduring ~\st.p = \cs~ \Bbegin
    ~\st.p \bcmeq \idle~ \Bend
  \end{align*}
\caption{Unit-B mutual exclusion specification}
\label{fig:unit-b:mutex}
\end{figure}

In addition to the \eventB constructs, the \unitb model features three
new ones: event coarse schedules, introduced by the keyword
\Bduring; event fine schedules, introduced by the keyword
\Bupon and event indices, denoted by the square brackets next
the the event names.  Intuitively, if the coarse schedules of an event
hold \emph{continually} and its fine schedules becomes true \emph{infinitely often}
then the event is executed infinitely often.

For all events (i.e., \request, \enter, \exit) , $p$ is an index
instead of an \eventB parameter (declared with the keyword
\Bany).  While a parameter is conceptually a value chosen
non-deterministically, an index suggests that there exists a distinct
version of the event, including a separate scheduling assumption, for
every one of the index's values. This allows us to prove individual
progress for each process $p$.
 
Event $\request$ is syntatically similar to its \eventB counter part.
Semantically, the difference is subtle but important.  In \eventB, if
$\request$ is the only enabled event, i.e. its guard is true and the
guard of every of other event is false, $\request$ will be taken
eventually. In \unitb, even if $\request$ is the only enabled event,
it might never occur. This is because $\request$ is not scheduled: it
features neither a coarse schedule (declared with \Bduring) nor a fine
schedule (declared with \Bupon).

Events $\enter$ and $\exit$ are scheduled events: $\enter$ has both a
coarse schedule and a fine schedule and $\exit$ has only a coarse
schedule.
In the case of $\exit$, when its coarse schedule is continually true,
i.e. some process is in its critical section and remains there, then
eventually $\exit$ is taken and $p$ exits its critical section. Event
$\enter$ is eventually taken if a process $p$ is waiting continually
(coarse schedule) and that infinitely often no process is in its
critical section (fine schedule). The fine schedule ensures that
$\enter$ occurs despite the other processes going in and out of their
critical sections.

The notion of schedule allows us to prove that certain events are
guaranteed to occur without having to reference the other events. This
is in contrast to \eventB where the only way to ensure that $\enter$ is
taken is to make sure that it is eventually the only event enabled.

The next step is to formulate the liveness requirement. In \unity
logic, on which \unitb is based, the absence of livelock (i.e.
\ref{intro:livelock}) is formulated as:
\[ \qexists{p}{}{\st.p = \waiting} \Wide \leadsto \qexists{p}{}{\st.p = \cs}~. \]
It reads ``whenever a process is waiting it eventually follows that a
process, possibly a different one, will gain access to its critical
section.''

Although the absence of livelock is an interesting property, it is too
weak to be useful; the goal of the processes is not to allow an
arbitrary other process to carry on with its work; it is rather the
goal of the mutual exclusion protocol to let the processes go about
their business unhindered, independently from each other. This means
that it is more important for the purpose of each process not to be left to
wait forever than any other property that has to do with the competing
processes. Therefore, we choose individual progress as the central
property to be proved. Its \unity formulation is:
\begin{align*}
 \tag{\Binv{prg1}}\label{intro:P1}
 \st.p = \waiting \Wide\leadsto \st.p = \cs
\end{align*}
The free variable $p$ is implicitly universally quantified over the whole formula.

In the process of proving \eqref{intro:P1}, we will discover that
another progress property is required. This is because the only way
that every process can safely have a turn in their critical section is
for no process to linger in theirs forever. We formulate it as
\eqref{intro:P0} and prove it first:
\begin{align*}
   \tag{\Binv{prg2}}\label{intro:P0}
   true \Wide\leadsto \qforall{p}{}{ \neg \st.p = \cs }
\end{align*}
It reads ``infinitely often, every process will be simultaneously out
of their critical section.'' In LTL, they are stated as:
\begin{align*}
   & \qforall{p}{}{ \square(~ \st.p = \waiting \1\implies \lozenge \st.p = \cs ~) }
     \tag{\Binv{prg1'}}
\\ & \square \lozenge \qforall{p}{}{\neg \st.p = \cs} 
     \tag{\Binv{prg2'}}
\end{align*}


The standard way of proving a liveness property in \unitb is to use
rules from \unity logic to transform the property into something that
is more easily proved using the events. 
The rules will be explained in more
details in Section~\ref{sec:contribution}.
For the sake of conciseness, we only sketch the intuition behind the
proofs of this example.

\paragraph{A sketched proof of \eqref{intro:P0}} As long as
$\qforall{p}{}{\neg \st.p = cs} $ does not hold, there exists a
process $p$ is in its critical section, i.e., the coarse schedule of
$\exit[p]$ is true.  Therefore, according to its scheduling assumption,
$\exit[p]$ eventually will, thus establishing $\qforall{p}{}{\neg \st.p
  = cs} $ in the process thanks to the mutual exclusion invariant,
\Binv{inv1}.

\paragraph{A sketched proof of \eqref{intro:P1}} Given a process $p$,
$\enter[p]$ is the only event that can establish $\st.p = cs$ and it
does so if $\st.p = waiting$ (its coarse schedule) holds continuously
and $\qforall{q}{p \neq q}{\neg \st.q = cs}$ (its fine schedule) is
true infinitely many times.  The latter is entailed by
\eqref{intro:P0} which we already proved.
The former condition is satisfied as soon as $\st.p = waiting$ (the
antecedent of \eqref{intro:P1}) is
established. 
.

It is very important to note that the index $p$ allows us to specify
the scheduling assumptions on a process-by-process basis. We can thus
assert and prove that every process will eventually acquire the lock \eqref{intro:P1},
a property otherwise known as \emph{starvation freedom}. This
property cannot be proved in \eventB.  In \eventB we can only prove
the weaker property \eqref{intro:livelock} that some arbitrary process will eventually
acquire the lock.

Traditionally, scheduling assumptions fall into two categories: weak
fairness and strong fairness. In this example, weak fairness (stating
that if a process $p$ is waiting and the lock is free
\emph{continually} then $p$ eventually takes the lock) is insufficient
to prove that process $p$ eventually takes the lock.  Normally, to
prove this property, the $\enter[p]$ event would be scheduled with
strong fairness (stating that if a process $p$ is waiting and the lock
is free \emph{infinitely often} then $p$ eventually holds the lock).
Using strong fairness, the coarse schedule (conditions required to
hold continually) and the fine schedule (conditions required to become
true infinitely often) would be wrapped in a single guard, thus
intertwining the reasoning about those two aspects. By decoupling
these orthogonal considerations, we can reason about process $p$
waiting separately from the lock becoming free. Moreover, during
refinement, this decoupling will allow us to trade freely between the
coarse and the fine schedules.  The distinction between coarse and
fine schedules and their relation to other scheduling assumptions are
explained further in Section~\ref{sec:contribution} and
Section~\ref{sec:second-refinement}.

The combination of the progress preserving refinement calculus with
the novel notions of coarse and fine schedules makes it possible in
\unitb to introduce liveness properties at any stage of a development
process.  Reasoning about both safety and liveness can be done at the
relevant abstractions. As a consequence, not only do we use liveness
requirements to rule out any design decision that would be too
conservative, but we also use them to guide us to the right design
decisions. As a result, liveness properties, in particular progress
properties, drive the development process.



\subsection{Contribution}
This paper features a formal semantics for \unitb models and their
properties, alongside an example of application of \unitb to a non-
trivial control problem.  The semantics is formulated in computation
calculus~\cite{Dijkstra:1998p1128}. We use it to formally prove the
soundness of the rules for reasoning about temporal properties and
refinement relationships in \unitb.

In the past, \unitb has been used to design a mutual exclusion
algorithm \cite{thesis/hudon2011} and a signal controller for a train
station \cite{ifm/Hudon/Hoang13}. This paper is an extended version of
the latter.  In addition to the contributions of
\cite{ifm/Hudon/Hoang13}, we (1) strengthen the separation between the
formal semantics and the proof obligations;  (2) present a new rule
permitting the reuse of progress properties without reproving them;
(3) formulate the refinement rules so as to allow the refinement of
events to be justified using only one rule; (4) elaborate the
individual refinement steps of the example with the design concerns
that guide it and the specific proof obligations;  (5) expand the
example with two refinement steps leading to the specification of a
controller and  (6) illustrate the use of inductive proofs of liveness
in the example.

\subsection{Structure} The rest of the paper is organised as follows.
In Section~\ref{sec:background}, we review Dijkstra's computation
calculus \cite{Dijkstra:1998p1128} which we used to formulate our
semantics and design our proofs.  We follow with a description of the
\unitb method in Section~\ref{sec:contribution}.  We demonstrate the
method and its refinement rules by developing a signal control system
in Section~\ref{sec:example}.  We summarise our work in
Section~\ref{sec:conclusion} including discussion about related work
and future work.



\section{Background: Computation Calculus}
\label{sec:background}

In this section, we give a brief introduction to computation calculus,
based on \cite{Dijkstra:1998p1128}.  This will be the basis for
defining the semantics of Unit-B models, defining the semantics of
temporal properties (both safety and liveness) and formulating the
proof of soundness of the \unitb refinement rules in
Section~\ref{sec:contribution}.

\begin{itemize}

\item In Section~\ref{sec:comp-pred}, we introduce the notion of
  computation predicates which can be manipulated algebraically.  We
  use them in Section~\ref{sec:contribution} to characterize the
  execution of \unitb models as well as their
  properties.

\item In Section~\ref{sec:state-predicates}, we introduce state
predicates, a special case of computation predicates which we use in
Section~\ref{sec:contribution} to formalize \unitb invariants,
progress and safety properties, events' guards and schedules.

\item In Section~\ref{sec:atomic-actions}, we introduce atomic
computation predicates, a special case of computation predicates which
we use in Section~\ref{sec:contribution} to formalize the meaning of
the events' actions.

\end{itemize}

Let $\State$ be the \emph{state space}: a non-empty set of ``states''.
Let $\Computation$ be the \emph{computation space}: a set of non-empty
(finite or infinite) sequences of states henceforth referred to as
``computations''.

\subsection{Computation Predicates}
\label{sec:comp-pred}

\begin{Definition}[Computation Predicates]  The set of computation predicates $\CPred$ 
  is defined as follows:
  \begin{align}
    \CPred &\WIDE= \Computation \rightarrow \BOOL~,
  \end{align}
  i.e., functions from computations to Booleans.
\end{Definition}

The standard Boolean operators of the predicate calculus are lifted,
i.e., extended to apply to $\CPred$. For example, assuming $s, t \in
\CPred$ and $\Comp \in \Computation$, we have,%
\footnote{%
  In this paper, we use $f.x$ to denote the result of applying a
  function $f$ to argument $x$.  Function application is
  left-associative, so $f.x.y$ is the same as $(f.x).y$.
}%
%
\begin{align}
  (s \implies t).\Comp  &\WIDE{\eqv}  (s.\Comp \implies
  t.\Comp) \label{eq:comp-impl}  \\
  \qforall{i}{}{s.i}.\Comp  &\WIDE{\eqv}
  \qforall{i}{}{s.i.\Comp}~. \label{eq:comp-forall}
\end{align}

The everywhere-operator quantifies universally over all
computations, i.e.,
\begin{align} 
\ew{s} & \WIDE\eqv \qforall{\Comp}{}{s.\Comp}~. \label{eq:comp-ew}
\end{align}
%
Whenever there are no risks of ambiguity, we
shall use $s = t$ as a shorthand for $\ew{s \eqv t}$ for computation
predicates $s, t$.

\begin{Postulate}
  \label{post:comp-pred-alg}
  $CPred$ is a predicate algebra.
\end{Postulate}
A consequence of \Post~\ref{post:comp-pred-alg} is that $\CPred$
satisfies all postulates for the predicate calculus as defined in
\cite{Dijkstra:1990:PCP:77545}.  In particular, $\ctrue$ (maps all
computations to $\True$) and $\cfalse$ (maps all computations to
$\False$) are the ``top'' and the ``bottom'' elements of the complete
Boolean lattice with the order $\ew{ \_ \implies \_}$ specified by
these postulates.  The lattice operations are denoted by various
Boolean operators including $\land, \lor, \neg, \implies$, etc.

The predicate algebra is extended with sequential composition as follows.%
\begin{Definition}[Sequential Composition]
  \begin{equation}
    \label{eq:comp-seq}
    \begin{array}{lcl}
    (s;t).\Comp  & \WIDE\eqv & (\size \Comp = \infty \land s.\Comp) \wide\lor\\
    & & \qexists{n}{n < \size \Comp}{s.(\Comp \take n\! + \!1) \land t.(\Comp
      \drop n)}
  \end{array}
  \end{equation}
where $\size$, $\take$ and $\drop$ denote sequence operations
`\emph{length}', `\emph{take}' and `\emph{drop}', respectively.
\end{Definition}
Intuitively, the sequential composition of $s$ and $t$ can be
understood as a program specification that requires $s$ to be run
first and then $t$ to be run as soon as $s$ terminates, if it does.
More specifically, a computation $\tau$ satisfies $s\,;t$ if either it
is an infinite computation satisfying $s$, or $\tau$ can be broken
into  a finite prefix $\tau \take n\! + \!1$ and a suffix $\tau \drop
n$ sharing state $\tau.n$ such that the prefix satisfies $s$ and the
suffix satisfies $t$.


In the course of reasoning using computation calculus, we make use of
the distinction between infinite (``eternal'') and finite
computations.  Two constants $\E, \F \in \CPred$ have been defined for
this purpose. 
\begin{Definition}[Eternal and Finite Computations] For any predicate $s$,%
  \begin{align}
    \E & \WIDE= \ctrue;\cfalse \label{eq:comp-eternal} \\
    \F & \WIDE= \neg \E \label{eq:comp-finite} \\
    \textrm{$s$ is eternal} & \WIDE\eqv \ew{s \wide\implies \E} \label{eq:comp-s-eternal} \\
    \textrm{$s$ is finite} & \WIDE\eqv \ew{s \wide\implies \F} \label{eq:comp-s-finite}
  \end{align}
\end{Definition}
An important property related to $\E$ (from~\cite{Dijkstra:1998p1128})
is that for any predicate $s$, we have
\begin{align}
  \label{eq:eternal-false}
  s;\cfalse &\WIDE= s \land \E~.
\end{align}

Given $\F$, the temporal ``eventually'' operator (i.e., $\diamondsuit$)
can be formulated as $\F;s$.  The ``always'' operator $\G$ is defined
as the dual of the ``eventually'' operator.
\begin{Definition}[Always Operator] For any predicate $s$,
  \begin{align}
    \G s &\WIDE= \neg(\F;\neg s)~.
  \end{align}
\end{Definition}
Important properties of $\G$ are that it is \emph{strengthening}~\eqref{eq:g-strengthen},
\emph{monotonic}~\eqref{eq:g-monotonic}, and it \emph{distributes over conjunction}~\eqref{eq:g-conjunctive}.  For any predicates $s$ and $t$, we have: 
\begin{align}
  \label{eq:g-strengthen}
  \ew{\G s &\WIDE\limp s}~, \\
  \label{eq:g-monotonic}
  \ew{s \wide\limp t} &\WIDE{\limp} \ew{\G s \wide\limp \G t}~, \\
  \label{eq:g-conjunctive}
  \G (s \wide\land t)  &\WIDE{ =} \G s \wide\land \G t~.
\end{align}

A useful technique that is frequently applied is to strip all the
outer $\G$ in some proofs, as illustrated in the following example.
For any predicates $s$, $t$, and $u$, we have
\begin{align}
  \label{eq:g-drop}
  \ew{s \land t \wide\limp u}  &\WIDE\limp \ew{\G s \wide\limp (\G t \limp \G u)}~.
\end{align}
The proof of \eqref{eq:g-drop} is as follows.
\begin{calculation}
  \ew{\G s \wide\limp (\G t \limp \G u)}
  \hint{=}{ shunting }
  \ew{\G s \land \G t \wide\limp \G u}
  \hint{=}{ $\G$ distributes through $\land$ \eqref{eq:g-conjunctive} }
  \ew{ \G (s \land t) \wide\limp \G u }
  \hint{\follows}{ monotonicity \eqref{eq:g-monotonic} }
  \ew{ s \land t \wide\limp u }
\end{calculation}

According to~\eqref{eq:g-drop}, whenever we need to prove a formula
of the form $\ew{\G s \wide\limp (\G t \limp \G u) }~,$  we can
reformulate it to strip the outer $\G$'s and manipulate $s$,$t$ and
$u$ on their own to simplify the proof.


\begin{Definition}[Persistence] For any predicate $s$,
  \begin{align}
    \textrm{$s$ is persistent} &\WIDE\eqv s = \G s~.
  \end{align}
\end{Definition}
A persistent predicate describes some repetitive mosaic.  If a
persistent s can be used to describe a computation $\tau$, s can also
be used to describe every suffix of $\tau$.  We borrow the following
facts related to the notion of persistence
from~\cite{Dijkstra:1998p1128}.  For all predicates $s$, $t$, and
persistent $u$, we have
\begin{align}
  \label{eq:g-persistent}
  &\textrm{$\G s$ is persistent, and} \\
  \label{eq:persistence}
  &\ew{ u \WIDE\limp (s\,;t \WIDE\equiv s\,;(t \land u))}~.
\end{align}
Consider some computation predicate $r$ where $\G r$ holds,
\eqref{eq:g-persistent} ensures that $\G r$ is persistent, and
\eqref{eq:persistence} (together with \eqref{eq:g-strengthen}) enables
us to insert $r$ after any sub-computation in a series of sequential
compositions. This is particularly useful when r is an invariant,
i.e., $r = p;\ctrue$, where $p$ is a \emph{state predicate} as defined
in the subsequent.

\subsection{State Predicates}
\label{sec:state-predicates}
A constant $\one$ is defined as the (left- and right-) neutral element for
sequential composition.
\begin{Definition}[Constant $\one$] For any computation $\tau$,
  \begin{align}
    \one.\tau &\WIDE{\eqv} \size \tau = 1
  \end{align}
\end{Definition}
An important property of $\one$ is that it is finite, i.e.,
\begin{align}
  \label{eq:1-finite}
  \ew{\one \limp \F}~.
\end{align}

In fact, $\one$ is the characteristic predicate of the state space.
Moreover, we choose not to distinguish between a single state and the
singleton computation consisting of that state, which allows us to identify
predicates of one state with the predicates that hold only for singleton
computations.  Let us denote the set of state predicates by 
$\SPred$.
\begin{Definition}[State Predicate] For any predicate $p$,
  \begin{align}
    p \in \SPred &\WIDE{\eqv} \ew {p \implies \one}~.
  \end{align}
\end{Definition}

A consequence of this definition is that $\SPred$ is also a complete
Boolean lattice with the order $\ew{ \_ \implies \_}$, with $\one$ and
$\cfalse$ being the ``top'' and ``bottom'' elements.  It inherits all
the lattice operators that it is closed under: conjunction,
disjunction, and existential quantification.  The other lattice
operations, i.e., negation and universal quantification, are defined by
restricting the corresponding operators on $\CPred$ to state
predicates.  We only use state predicate negation in this paper.
\begin{Definition}[State predicate negation $\spneg$] For any state
  predicate $p$, 
  \begin{align}
    \spneg p &\WIDE= \neg p \land \one~.
  \end{align}
\end{Definition}

For a state predicate $p$, the set of computations with the initial
state satisfying $p$ is captured by $p\,;\ctrue$: the weakest such 
predicate.  A special notation $\initially : \SPred \rightarrow \CPred$ 
is introduced to denote this predicate.
\begin{Definition}[Initially Operator $\initially$] For any state predicate $p$,
  \begin{align}
    \initially p \WIDE= p\,;\ctrue~.
  \end{align}
\end{Definition}

This entails the validity of the following rule, which we will use
anonymously in the rest of the paper: for $p, q$ two \emph{state
  predicates}, 
\begin{align}
  p \, ; q \WIDE= p \land q~.
\end{align}
Another common rule related to state predicate is the
\emph{state restriction} rule allowing to trade $\land$ and
$\initially$ for $;$ and vice versa.  Given any predicate $s$ and any state
predicate $p$, we have
\begin{align}
  \label{eq:state-restriction}
  p\,;s &\WIDE= s \land \initially p~.
\end{align}

\subsection{Atomic Actions}
\label{sec:atomic-actions}
An important operator in LTL is the ``next-time operator''.   This is
captured in computation calculus by the notion of atomic computations:
computations of length 2.  A constant $\X \in \CPred$ is defined for
this purpose.
\begin{Definition}[Atomic Actions] For any computation $\tau$ and
  predicate $a$,
  \begin{align}
    \X.\tau  & \WIDE\eqv  \size \tau = 2  \label{eq:comp-next} \\
    \textrm{$a$ is an atomic action} & \WIDE\eqv  \ew {a \implies \X} 
  \end{align}
\end{Definition}
Given the above definition, the ``next'' operator can be expressed
as $\X\,;s$ for arbitrary computation $s$.
An important property for $\X$ is that it is finite, i.e.,
\begin{equation}
  \label{eq:X-finite}
  \ew{\X \limp \F}~.
\end{equation}



\section{The Unit-B Method}
\label{sec:contribution}
\newBmch[Mch]{M}
\newBevt[evt]{e}

This section presents our contribution: the \unitb method. It is
inspired by \eventB~\cite{DBLP:books/daglib/0024570} and
\unity~\cite{DBLP:books/daglib/0067338}.

Similar to \eventB, \unitb is aimed at the design of software systems
by stepwise refinement, where each step is verified via the
application of correctness preserving refinement rules.  It differs
from \eventB by its capability for reasoning about progress properties
and by its refinement order which preserves liveness properties.  It
also differs from \unity by unifying the notions of programs and
specifications, allowing stepwise refinement of programs from abstract
models.

\begin{itemize}

\item In Section~\ref{sec:syntax}, we briefly review the syntax of
  \unitb models (which has been informally introduced earlier in
  Section~\ref{sec:unitb-model}).

\item In Section~\ref{sec:semantics}, we describe the semantics of a
\unitb model $\Mch$. We characterize the set of executions of $\Mch$
by a computation predicate $ex.\Mch$ which is the conjunction of a
safety and a liveness component. We provide proof rules for invariant
preservation and \emph{unless properties}.
We also show the that the proof rules are sound with
respect to the semantics.

\item In Section~\ref{sec:progress-properties}, we provide proof rules
for progress properties and prove their soundness.

\item In Section~\ref{sec:refinement}, we provide refinement rules and
prove their soundness with respect to the semantics.

\end{itemize}

A more extensive discussion of the soundness of the proof rules of \unitb
is presented in \cite{thesis/hudon2011}.

\subsection{Syntax} 
\label{sec:syntax}
Similar to \eventB, a \unitb system is modelled by a transition system,
where the state space is captured by variables $\var$ and the transitions are
modelled by guarded events.  Furthermore, \unitb has additional 
assumptions on how the events should be scheduled.  Using
an \eventB-similar syntax, a \unitb event has the following form:
\begin{equation}
  \ubeventinlineidx{\evt}{\idx}{\csched.\idx.\var}{\fsched.\idx.\var}{}{\guard.\idx.\var}{\assignment.\idx.\var.\var'}
  \label{eq:ubevent}
\end{equation}
where $\idx$ are the event's \emph{indices},
$\guard$ is the event's \emph{guard}, $\csched$ is the event's \emph{coarse schedule},
$\fsched$ is the event's \emph{fine schedule}, and $\assignment$ is
the event's
\emph{action} changing state variables $\var$.  The action is usually
made up of several \emph{assignments}, either deterministic ($\bcmeq$)
or non-deterministic ($\bcmsuch$ or $\bcmin$).
An event $\evt$ with indices $\idx$ stands for multiple events.  Each
corresponds to several non-indexed events $\evt.\idx$, one for each
possible value of the indices $\idx$.  Here $\guard$, $\csched$,
$\fsched$ are state predicates.  An event $\evt$ is said to be enabled
when its guard $\guard$ holds.  The scheduling assumption of the event
is specified by $\csched$ and $\fsched$ as follows: if
\emph{$\csched$ holds continually} and \emph{$\fsched$ becomes true
  infinitely often} then event $\evt$ is carried out infinitely often.
An event without any scheduling assumption will have its coarse
schedule $\csched$ equal to $\cfalse$.  An event having only the
coarse schedule $\csched$ will have the fine schedule to be $\one$.
Vice versa, an event having only the fine schedule $\fsched$ will have
the coarse schedule to be $\one$.%

In addition to the variables and the events, a model has an
initialisation state predicate \init constraining the initial value of
the state variables.
All computations of a model start from a state satisfying the
initialisation and are such that, at every step, either one of its
enabled events occurs or the state is unchanged, and each computation
satisfies the scheduling assumptions of all events.

\subsection{Semantics} 
\label{sec:semantics}

In the following, we use computation calculus to
give the formal semantics of \unitb models.  Let $\Mch$ be a \unitb model
containing a set of events of the form~\eqref{eq:ubevent} and an
initialisation predicate $\init$.  
Since the action of the event can be described by a before-after
predicate $\assignment.\idx.\var.\var'$, it corresponds to an atomic action
\begin{align}
  \Action.\idx &\WIDE= \qforall{x}{}{\initially (x = \var)
    \2\limp \X \, ; \assignment.\idx.x.\var}~.
\end{align}
In the above, the quantified variable $x$ is introduced to capture the
before value of $\var$, hence allows us to relate the pre-state and
the post-state using the state predicate $\assignment.\idx.x.\var$
applied to the post-state
(as indicated by $\X\,;\underline{\hspace{.2cm}}$).
%
Given that an event $\evt.\idx$ can only be carried out when it is
enabled, we formulate the effect of each event execution as follows: 
\begin{align}
  \action.(\evt.\idx) &\WIDE= \guard.\idx\,\1;\,\Action.\idx~~.
\end{align}


The semantics of \Mch is given by a computation predicate $\execution.\Mch$
which is a conjunction of a ``safety part'' $\safety.\Mch$ and a
``liveness part'' $\liveness.\Mch$ (both to be defined later), i.e., 
\begin{align}
  \ew{\execution.\Mch &\WIDE{\eqv} \safety.\Mch \wide\land \liveness.\Mch}~.\label{eq:execution}
\end{align}

\begin{Definition}
  A property $s$ is satisfied by \Mch, denoted $\Mch \models s$, if
  the property is implied by $\execution.\Mch$.  
  \begin{equation}
    \Mch \wide\models s \WIDE{\textrm{if and only if}} \ew{\execution.\Mch \wide\limp s}~.\label{eq:property}
  \end{equation}
\end{Definition}
We use $\Mch \vdash s$ to denote that \emph{$\Mch \models
s$ is provable}.

Properties of \unitb models are captured by two types of properties:
\emph{safety} and \emph{progress} (liveness).

\subsubsection{Safety} Below, we define the general form of one step
of execution of model \Mch, i.e., $\step.\Mch$, and the \emph{safety}
constraints $\safety.\Mch$ on its
complete computations.
\begin{align}
  \ew{\step.\Mch  &\WIDE{\eqv} \qexists{\evt, \idx}{\evt.\idx \in \Mch}{\action.(\evt.\idx)} \,\lor\, \Skip} \label{eq:step} \\
  \label{eq:safety}
  \ew{\safety.\Mch  &\WIDE{\eqv}  \initially \init \land
    \G(\step.\Mch \, ; \,\ctrue)}
\end{align}
%
where $\Skip$ is a special unscheduled event that is a part of every 
model. Its guard is true and its effect is to leave all the variables of 
model unchanged. Since the variables of the current model may be 
only one
of the components of the state space --- the other components being the 
variables of the models that may refine the current one --- $\Skip$ 
makes no commitment about the final value of those components; that is
to say that they will be changed non-deterministically without constraints. 

Safety properties of the model are captured by \emph{invariance} properties
(also called \emph{invariants}) and by \emph{unless} properties. 

\paragraph{Invariance properties}
An invariant $I.v$ is a state-property that holds at every reachable state of the model.
If $I.v$ is an invariant of $\Mch$, in all executions of $\Mch$, $I.v$ holds forever:
\begin{align}
  \ew{\execution.\Mch \WIDE{\limp} \G \initially \! I}~.
\end{align}
In particular, we rely solely on the safety part of the model to prove
invariance properties, i.e., we prove $\ew{\safety.\Mch \wide{\limp} \G
  \initially \! I}$.  This leads to the well-known invariance principle.
\begin{equation}
  \small
  \proofrule{%
    \phantom{\Mch} \WIDE\vdash \init.\var \wide\limp I.\var \\
    \phantom{\Mch} \WIDE\vdash I.\var \land \guard.\idx.\var \land
    \assignment.\idx.\var.\var' \wide\limp I.\var' \quad\textrm{(for all
      event $\evt.\idx$)}
  }
  {%
    \Mch \WIDE\vdash \G \initially I
  }
  \tag{INV}
\end{equation}
Invariance properties are important for reasoning about the
correctness of the models since they give an (over-)approximation of
the set of reachable states.  This makes it possible to use invariance
properties as additional assumptions in proofs for other properties
(often as a consequence of applying \eqref{eq:g-persistent} and
\eqref{eq:persistence}).  For example, we can propagate a state
predicate $I$ to the middle of a sequential composition $s;t$ as
follows: under the assumption that $I$ holds forever, i.e., $\G
\initially I$, either because it is an invariant or for other reasons,
for any predicates $s$ and $t$, we have
\begin{equation}
  \label{eq:invariant-middle}
  s\,;t \wide = s\,;I\,;t
\end{equation}
The proof of \eqref{eq:invariant-middle} is as follows.
\begin{calculation}
  s;t
  \hint{=}{ $\G \initially I$ (persistent) with
  persistence rule \eqref{eq:persistence} }
  s;(t \land \G \initially I)
  \hint{=}{ $\G$ is strengthening \eqref{eq:g-strengthen} }
  s;(t \land \initially I \land \G \initially I)
  \hint{=}{ $\G \initially I$ (persistent) with
  persistence rule \eqref{eq:persistence} }
  s;(t \land \initially I)
  \hint{=}{ state restriction \eqref{eq:state-restriction} }
  s;I;t
\end{calculation}

In the subsequent, we assume that model \Mch has an invariant $I.v$.

\paragraph{Unless properties}
The other important class of safety properties is defined by the
\emph{unless} operator $\un$.
\begin{Definition}[$\un$ operator]  For any state predicates $p$,
  $q$,
  \begin{equation}
    \label{eq:un-def}
    \ew{(p \un q) \WIDE{\eqv} \G (\initially p \wide\limp (\G \initially \! p)\,;(\one \lor \X)\,;\initially q)}
  \end{equation}
\end{Definition}
Informally, $p \un q$ is a safety property stating that if condition
$p$ holds then it will hold continuously unless $q$ becomes true.
The formula $(\one \lor \X)$ is used in \eqref{eq:un-def} to allow
the last state where $p$ holds and the state where $q$ first holds to 
either be the same state or to immediately follow one another.

The following theorem is used for proving that a \unitb model
satisfies an unless property.
\begin{Theorem}[Unless rule]
  \label{thm:unless}
  Consider a model \Mch with invariant $I$ and an unless property
  $p.\var \un q.\var$. We have 
  \[ \ew{\execution.\Mch \wide\limp p \un
  q}\] 
  if for every event  \evt and index value \idx with $\evt.\idx
  \in \Mch$,
  \begin{align}
      \ew{
          (I \land p \, \land \! \spneg q) ;\action.(\evt.\idx) 
          \2\limp 
          \X ; \! (p \lor q) 
      }
    ~.
    \label{eq:unless}
  \end{align}
\end{Theorem}
\begin{proof}[Sketch]
  Condition \eqref{eq:unless} ensures that every event $\evt.\idx$ of
  \Mch either maintains $p$ or establishes $q$. By induction, we can
  see that the only way for $p$ to become false after a state where it
  was true is that either $q$ becomes true or that it was already
  true. The full proof can be found in \cite[Section~2.0.1]{thesis/hudon2011}
  \qed
\end{proof}

It follows from \Thm~\ref{thm:unless} 
that the following proof rule can be used to prove unless properties.
\begin{equation}\small
  \label{eq:un-rule}
  \proofrule{
    \phantom{\Mch} \WIDE\vdash \begin{array}[c]{cl}
      & p.\var \wide\land \neg \, q.\var
      \wide\land I.\var \wide\land \\
      & \guard.\idx.\var  
      \wide\land \assignment.\idx.\var.\var' \\
      \wide\implies & p.\var' \1\lor  q.\var' \quad\quad\quad\quad\textrm{(for all event $\evt.\idx$)}
    \end{array}
  } { \Mch \WIDE\vdash p \1\un q}
  \tag{UN}
\end{equation}
The antecedent of \eqref{eq:un-rule} has the interesting peculiarity
that it does not include either the fine or the coarse schedule of
event \evt.

\subsubsection{Liveness}
For each event of the form~\eqref{eq:ubevent}, its schedule
$\schedule.(\evt.\idx)$ is formulated as follows, where $\csched$ and
$\fsched$ are the event's coarse and fine schedule, respectively:
\begin{equation}
  \label{eq:schedule}
  \left[\schedule.(\evt.\idx) \Wide{\eqv} \G \left(
      \begin{array}[c]{ll}
        & ~~ \G \initially \!\csched.i
        \,\land\, \G \F\,;\initially \fsched.i  \\
        & \limp\\
        & ~~ \F\,;\fsched.i\,;\action.(\evt.\idx)\,;\ctrue
    \end{array}
  \right)
  \right]~.
  \tag{SCH}
\end{equation}
Intuitively, \eqref{eq:schedule} states that if the coarse schedule
$\csched$ holds \emph{continually}, i.e., $\G \initially\csched$
and the fine schedule $\fsched$ becomes true \emph{infinitely often}, i.e.,
$\G \F;\initially\fsched$, then eventually $\evt.\idx$ occurs at a
point where $\fsched$ holds, i.e., $\F;\fsched;\action.(\evt.\idx)$.
To ensure that the event $\evt.\idx$ only occurs when its guard
$\guard.\idx$ holds, we require the following \emph{feasibility} condition:
\begin{equation}
  \label{eq:fis}  
  I.\var \land \csched.\idx.\var \land \fsched.\idx.\var \Wide\limp \guard.\idx.\var
  \tag{SCH\_FIS}
\end{equation}
In absence of this condition, the (coarse or fine) schedule may
be continuously in contradiction with the guard: while the scheduling 
constraint \eqref{eq:schedule} states that all valid computation will
include occurrences of the event, the safety constraint 
\eqref{eq:safety} states that, under the same conditions, the event 
will not happen. It follows that no traces satisfy the two constraints
and the system cannot be implemented.

Our coarse and fine schedules are a generalisation of the standard
weak-fairness and strong-fairness assumptions.  The standard
\emph{weak-fairness} assumption for event $\evt.\idx$ (stating that if
$\evt.\idx$ is enabled continually then eventually it will be
taken) can be formulated by using $\csched = \guard$ and $\fsched =
\one$.  Similarly, the standard \emph{strong-fairness} assumption for
$\evt.\idx$ (stating that if $\evt.\idx$ is enabled infinitely often then
eventually it will be taken) can be formulated by using $\csched =
\one$ and $\fsched = \guard$.
\begin{align*}
  \ew{ \wf.(\evt.\idx)  &\WIDE\equiv \G (\G \bullet \guard.\idx \1\implies
    \F;\action.(\evt.\idx);\ctrue) } \\
  \ew{ \strf.(\evt.\idx)  &\WIDE\equiv \G (\G \F;\bullet \guard.\idx \1\implies \F;\action.(\evt.\idx);\ctrue) }
\end{align*}
\indent
Instead of categorizing \unitb events between weakly fair and strongly
fair, our generalization allows us to have a little of both in every
event. Strong fairness is often a nice abstraction of scheduling magic
happening under the hood but it is necessary to refine it away in
order to implement it. Our generalization facilitates this by making
the transition between strong fairness to weak fairness smoother. 

In Section~\ref{sec:cf:sched:vs:w:str:fair}, we provide a
methodological comparison between weak and strong fairness on one hand
and coarse and fine schedules on the other hand in the context of the
main example. Furthermore, in Section~\ref{sec:second-refinement} we
discuss the heuristics justifying the choice of coarse and fine
schedules of events.

The liveness part of the model is the conjunction of the schedules for
its events, i.e.,
\begin{equation}
  \label{eq:liveness}
  \ew{\liveness.\Mch \Wide{\eqv} 
     \qforall{\evt, \idx}{\evt.\idx \in \Mch}{\schedule.(\evt.\idx)}}~
\end{equation}
The summary of the \unitb modelling notation is showed in 
Figure~\ref{fig:unitb}.
\begin{figure}[!htbp]
  \centering
  $%
  \ubeventinlineidx{\evt}{\idx}{\csched.\idx.\var}%
  {\fsched.\idx.\var}{}{\guard.\idx.\var}%
  {\assignment.\idx.\var.\var'}%
  $
    
  \begin{align*}
    \ew{&\execution.\Mch & \WIDE\eqv &\safety.\Mch \wide\land
      \liveness.\Mch}\\
    \ew{&\safety.\Mch & \WIDE\eqv &\initially \init \wide\land
      \G(\step.\Mch;\ctrue)} \\
    \ew{&\step.\Mch & \WIDE\eqv & \qexists{\evt,\idx}{\evt.\idx \in
        \Mch}{\action.(\evt.\idx)} \wide\lor \Skip} \\
    \ew{& \action.(\evt.\idx) &\WIDE\eqv & \guard.\idx.\,;\,\Action.\idx}\\
    \ew{& \Action.\idx.\prm &\WIDE\eqv & \qforall{x}{}{\initially (x = \var)
        \wide\limp \X \, ; \assignment.\idx.x.\var}}\\
    \ew{& \liveness.\Mch &\WIDE{\eqv} &\qforall{\evt, \idx}{\evt.\idx
        \in \Mch}{\schedule.(\evt.\idx)}}\\
    \ew{&\schedule.(\evt.\idx) &\WIDE{\eqv} &
      \G(
      \G \initially \!\csched.i
      \,\land\, \G \F\,;\initially \fsched.i
      \wide\limp
      \F\,;\fsched.i\,;\action.(\evt.\idx)\,;\ctrue
      )
    } \\
    & \Mch  \models s & \WIDE{\textrm{iff}} & \ew{\execution.\Mch \wide\limp s}
  \end{align*}

  \caption{Summary of \unitb}
  \label{fig:unitb}
\end{figure}

\subsection{Progress Properties}
\label{sec:progress-properties}
Progress properties are of the form $p \leadsto q$, where $\leadsto$
is the leads-to operator. They state that every state satisfying
predicate $p$ is eventually followed by by a state satisfying $q$.
\begin{Definition}[$\leadsto$ operator] For any state predicates $p$,
  $q$,
  \begin{align}
    \label{eq:leadsto}
    \ew{(p \leadsto q) &\WIDE{\eqv} \G (\initially p \wide{\limp} \F \, ; \initially  q)}
  \end{align}
\end{Definition}
In the case where $p$ and $q$ contain free variables, i.e., variables
not belonging to the state space, $p \leadsto q$ is understood
implicitly as
\[ \qforall{x}{}{p \leadsto q} \]
with $x$ the tuple of all the free variables appearing in either $p$ or $q$. 
The same principle is applied to unless, transient and falsifies properties,
the last two are introduced later in this section.



A special kind of progress properties is captured by the
\emph{transient operator}. Transient property $\tr p$ states  that
whenever predicate $p$ holds, it is eventually falsified. Transient
properties are especially useful for creating a bridge between leads-
to properties and the events that effect them. That bridge is
completed by the $\falsifies$ operator which we introduce later in
this section.

\begin{Definition}[$\tr$ operator] For any state predicate $p$,
  \label{def:transient}
  \begin{align}
    \tr p  \WIDE= p \leadsto \spneg p 
    \WIDE= \one \leadsto \spneg p 
    \WIDE= \G \F;\initially \spneg p
    \label{eq:transient}
  \end{align}
\end{Definition}

The properties of $\leadsto$ and $\tr$ that we will use in
this paper are as follows. For any state
predicates $p$, $q$, and $r$, we have:
\begin{gather}
  \ew{ \G \initially\! (p \limp q) \WIDE{\limp} (p \leadsto q)}%
  \tag{Implication}\label{eq:implication} \\
  \ew{ (p \leadsto q) \wide\land (q \leadsto r) \WIDE{\limp} (p \leadsto r)
  }\tag{Transitivity}\label{eq:transitivity} \\
  \label{eq:37} 
  \ew{ (p \leadsto q)  \WIDE{\eqv}  (p \land \! \spneg q \wide{\leadsto} q)
  }\tag{Split-Off-Skip}\\
  \ew{(p \wide\un q) \wide\land (\tr p \land \spneg q) \2\limp (p
    \leadsto q)}
  \tag{Ensure} \label{eq:ensure}\\[1ex]
  \left[
    \begin{array}{rl}
      & (p \wide\land v = M \Wide\leadsto (p \wide\land v < M) \wide\lor q) \\ 
      \limp~ & (p \wide\leadsto q) 
    \end{array}
  \right] \tag{Induction}\label{eq:induction} \\[1ex]
  \label{eq:psp} 
  \left[
  \begin{array}{rl} 
  	& (p \wide\leadsto q)  \Wide\land (r \wide\un b) \\
	\implies~ &  (p \land r \Wide{\leadsto} (q \land r) \lor b)
  \end{array}
  \right] \tag{PSP} 
\end{gather}

Above, in the induction rule, $M$ is a free variable and $v$ is the
variant, an expression involving some state variables. The name of the
\eqref{eq:psp} rule  stands for \emph{Progress, Safety, Progress}.
Except for \eqref{eq:37}, the above rules are taken from
\cite{DBLP:books/daglib/0067338}.


We prove progress properties by relating them to
the events of the model with $\falsifies$ properties. We can establish $\tr p$ by choosing an event
$\evt$ of the model and proving $\evt \falsifies p$, i.e., if $p$
holds continually \evt is eventually taken and whenever \evt is
executed in a state where $p$ holds it falsifies $p$.

\begin{Definition}[$\falsifies$ operator] For any state predicate $p$
  \label{def:falsifies}
  and any event as follows.:
  \[
  \ubeventinlineidx{\evt}{\idx}
    {\csched.\idx.\var}
    {\fsched.\idx.\var}{}
    {\guard.\idx.\var}
    {\assignment.\idx.\var.\var'}
  \]
  Event \evt with (actual) index \idx \emph{falsifies} property $p$
  (denoted as $\evt.\idx \falsifies p$) if under condition $p$,
  $\evt.\idx$ negates $p$ in one step \eqref{eq:NEG}, the coarse schedule \csched
  is enabled \eqref{eq:C-EN}, and the fine schedule \fsched is
  eventually enabled \eqref{eq:F-EN}.
  \begin{align}
  \ew{ \evt.\idx \falsifies p &\WIDE\equiv \eqref{eq:NEG}
    \wide\land \eqref{eq:C-EN} \wide\land \eqref{eq:F-EN}  }~,
  \end{align}
  where
  \begin{gather}
    \G ~\left(
     (p \land \csched \land \fsched); ~ \action.(\evt.\idx) \1; \true
    \wide\limp \X\,; \initially \spneg p
     \right)~,
     \label{eq:NEG}
    \tag{NEG} \\
    \G ~\initially\! (p \wide{\limp} \csched)~,
    \tag{C\_EN} 
    \label{eq:C-EN}
    \\
    \label{eq:F-EN}
    \tag{F\_EN} 
    p \land \csched \WIDE{\leadsto} \fsched~.
\end{gather}

%
%
\end{Definition}
Property $\evt.\idx \falsifies p$ states that, when state predicate $p$ 
holds, if it is not falsified by events other than $\evt.\idx$, 
$\evt.\idx$ will eventually occur and falsify $p$.

Given the definition of $\falsifies$, we have the following proof
rule (taking into account the invariant $I.\var$).
\begin{equation}
  \small
  \label{eq:falsifies}
  \proofrule{%
    \phantom{\Mch} \WIDE\vdash I.\var \land p.\var \wide\land  \csched.\idx.\var \wide\land  \fsched.\idx.\var
    \wide\land  \assignment.\idx.\var.\var'
    \Wide\limp \neg p.\var'  \\
    \phantom{\Mch} \WIDE\vdash I.\var \land p.\var \wide\limp \csched.\var \\
    \Mch \WIDE\vdash p.\var \land \csched.\idx.\var \wide\leadsto \fsched.\idx.\var
  }
  {
    \Mch \WIDE\vdash \evt.\idx \falsifies p.\var
  }
  \tag{FLS}
\end{equation}



%
%

The $\falsifies$ properties are the main tool for linking the model
and the progress properties in \unitb.  The attractiveness of such
properties is that we can \emph{implement} them using a single event.
In the case of events without a fine schedule (i.e., $\fsched$ is
$\one$), which is the most common one, the last condition
\eqref{eq:F-EN} becomes trivial and can be omitted.

\begin{Theorem}[Transient rule]
  \label{thm:tr-rule}
  Consider state predicate $p$ and a model \Mch contains event $\evt$.
  \[
  \ubeventinlineidx{\evt}{\idx}{\csched.\idx.\var}{\fsched.\idx.\var}{}{\guard.\idx.\var}{\assignment.\idx.\var.\var'}
  \]
  Given an (actual) index \idx, we have
  \begin{center}
    $\ew{\execution.\Mch \wide\limp \tr p}$ \Wide{if} $\ew{\execution.\Mch
      \wide\limp \evt.\idx \falsifies p}$~.
  \end{center}
\end{Theorem}
\begin{proof}
  Unfolding the definitions of $\tr$ and $\falsifies$, we prove $\G
  \F;\initially \spneg p$ under the assumptions \eqref{eq:NEG},
  \eqref{eq:C-EN} and \eqref{eq:F-EN}.  Moreover, since \evt is an
  event in \Mch, we have $\ew{\execution.\Mch \wide\limp
    \schedule.(\evt.\idx)}$~.  Therefore we have
  $\schedule.(\evt.\idx)$ as an additional assumption.

  Dropping the outer $\G$ in the goal and in the assumptions (similar
  to \eqref{eq:g-drop}), our goal becomes $\F;\initially \spneg p$.
  Additionally, since $\ew{ \neg s \limp s \wide{\eqv} s }$ for any
  computation predicate $s$, we discharge our obligation by
  strengthening $\F\; ; \, \initially \! \spneg p$ to its negation,
  $\G \initially \! p$.

  \begin{calculation}
    \F \, ; \initially\! \spneg p
    \hint{\follows}{ $\ew{ \F\,;\X \limp \F}$, aiming for \eqref{eq:NEG} }
    \F \, ; \X \, ; \initially \! \spneg p
    \hint{\follows}{ \eqref{eq:NEG} }
    	\F\, ; (p \land c \land f) \, ; \action.(\evt.\idx) \, ; \ctrue
    \hint{\follows}{ property of $\G$ \eqref{eq:invariant-middle} }
	\F ;\! f ; \! \action.(\evt.\idx) ; \! \ctrue \wide{\land} \G \initially \! c \wide{\land} \G \initially \! p
    \hint{\follows}{ $\schedule.(\evt.\idx)$ and definition \eqref{eq:schedule} }
    \G  \F ; \! \initially f \wide{\land} \G \initially\! c
    \wide{\land} \G \initially \! p
    \hint{=}{ $\G$ distributes through $\land$ \eqref{eq:g-conjunctive}}
    \G (~\F ; \! \initially f \wide{\land} \initially\! (c\land  p)~)
    \hint{=}{ \eqref{eq:F-EN} }
    \G \initially\! (c\land  p)
    \hint{=}{ \eqref{eq:C-EN} }
    	\G \initially \! p
    \end{calculation}
\qed
\end{proof}
\Thm~\ref{thm:tr-rule} corresponds to the following proof rule.
  \begin{equation}
    \proofrule{%
      \Mch \Wide\vdash \evt.\idx \falsifies p
    }{
      \Mch \Wide\vdash \tr p
    }
    \label{eq:tr-rule}
    \tag{TRS}
  \end{equation}


\subsection{Refinement}
\label{sec:refinement}
\newBmch[cncMch]{N}

In this section, we develop rules for refining \unitb models such that
safety and liveness properties are preserved.  Consider models \absMch
and \cncMch. Refinement, denoted by $\Mch \sqsubseteq \cncMch$ is defined by:
\begin{align}
  \label{eq:ref}
  \Mch \1\sqsubseteq \cncMch  \3\eqv 
  \ew{\execution.\Mch &\Wide\follows \execution.\cncMch}~.
  \tag{REF}
\end{align}
We call \absMch the abstract model and \cncMch the concrete model.
As a result of this definition, any property of \absMch is also
satisfied by \cncMch.  Similarly to \eventB, refinement is considered
in \unitb on a per event basis.  Each abstract event \absevt is
\emph{refined} by a concrete event \cncevt.
\begin{gather}
  \ubeventinlineidx{\absevt}{\absidx}%
  {\abscsched.\absidx.\var}{\absfsched.\absidx.\var}%
  {}%
  {\absguard.\absidx.\var}{\assignment.\absidx.\var.\var'} \\
  \ubeventinlineidx{\cncevt}{\cncidx}%
  {\cnccsched.\cncidx.\var}{\cncfsched.\cncidx.\var}%
  {}%
  {\cncguard.\cncidx.\var}{\assignment.\cncidx.\var.\var'}
\end{gather}
We say that \cncevt refines \absevt if
\begin{gather}
  \qforall{\cncidx}{}{%
    \left(
      \begin{array}[c]{l}
        \exists \absidx :: [\,\execution.\cncMch \wide\limp \\
        \quad (\action.(\cncevt.\cncidx) \limp
        \action.(\absevt.\absidx)) \, ]
      \end{array}
    \right)
  }
  \tag{EVT\_SAFE} 
  \label{eq:evt-safety}\\
  \qforall{\absidx}{}{%
    \left(
      \begin{array}[c]{l}
        \exists \cncidx::[ \, \execution.\cncMch \wide\limp \\
        \quad \! (\schedule.(\cncevt.\cncidx) \limp
        \schedule.(\absevt.\absidx)) \, ]
    \end{array}
      \right)
  }
  \tag{EVT\_LIVE} 
  \label{eq:evt-live}
\end{gather}

The proof that \cncMch refines \absMch (i.e., \eqref{eq:ref}) given
conditions \eqref{eq:evt-safety} and \eqref{eq:evt-live} is left out.
A special case of event refinement is when the concrete event $\cncevt$
is a new event.  In this case, we prove that \cncevt is the refinement of
the special \Skip event which is unscheduled and does not change any
variables of the abstract model.

Condition \eqref{eq:evt-safety} leads to similar proof obligations in
\eventB such as \emph{guard strengthening} and \emph{simulation}.  We
focus here on expanding the condition \eqref{eq:evt-live}.

We consider two cases for event refinement: (1) the abstract and
concrete events have the same indices, and (2) the indices are removed
from the concrete event.
\begin{Theorem}[Retaining Events' Indices]  Consider events \absevt and
  \cncevt as follows.
  \label{thm:schedule-ref}
  \begin{gather}
    \label{eq:absevt1}
    \ubeventinlineidx{\absevt}{\idx}%
    {\abscsched.\idx.\var}{\absfsched.\idx.\var}%
    {}%
    {\absguard.\idx.\var}{\assignment.\idx.\var.\var'} \\
    \label{eq:cncevt1}
    \ubeventinlineidx{\cncevt}{\idx}%
    {\cnccsched.\idx.\var}{\cncfsched.\idx.\var}%
    {}%
    {\cncguard.\idx.\var}{\assignment.\idx.\var.\var'}
  \end{gather}
  Assume \eqref{eq:evt-safety} have been proved for \absevt and
  \cncevt, i.e.,
  \begin{gather}
    \action.(\cncevt.\idx) \Wide\limp
    \action.(\absevt.\idx)~. \label{ref:safety}
  \end{gather}
  Given
  \begin{gather}
    \abscsched \land \absfsched
    \WIDE\leadsto \cnccsched
    \tag{C\_FLW}\label{ref:c-flw} \\
    \label{ref:c-stb}
    \cnccsched \WIDE\un \spneg \abscsched \tag{C\_STB} \\
    \abscsched \land \absfsched
    \WIDE\leadsto \cncfsched\tag{F\_FLW}\label{ref:f-flw} \\
    \G \initially(\cnccsched \land \cncfsched \wide \limp
    \absfsched)\tag{F\_STR} \label{ref:f-str}
  \end{gather}
  then 
  \begin{align}
    \schedule.(\cncevt.\idx) \wide\limp \schedule.(\absevt.\idx)\label{ref:sched}
  \end{align}
\end{Theorem}
\begin{proof}

  We first prove that the left-hand side of
  $\schedule.(\absevt.\idx)$, i.e., $\G \initially \abscsched \land \G
  \F;\initially \absfsched$ eventually leads to the left-hand side of
  $\schedule.(\cncevt.\idx)$, i.e., $\G \initially \cnccsched \land \G
  \F;\initially \cncfsched$.
  \begin{align}
    \G (\G \initially \abscsched \land \G \F;\initially \absfsched &\Wide\limp \F;
    (\G \initially \cnccsched \land \G \F;\initially \cncfsched)) \label{eq:1}
  \end{align}
  Dropping the outer $G$ from \eqref{eq:1}, we start the proof from
  $\G \F;\initially \absfsched$ with assumption $\G \initially
  \abscsched$.
  \begin{calculation}
    \G \F;\initially \absfsched
    \hint{\limp}{ $\G \initially \abscsched$ }
    \G \F;\initially (\abscsched \land
    \absfsched)
    \hint{\limp}{ \eqref{ref:c-flw} and \eqref{ref:f-flw} }
    \G \F;\initially \cnccsched \Wide\land \G \F;\initially \cncfsched
    \hint{\limp}{ \eqref{ref:c-stb} and definition of $\un$ \eqref{eq:un-def} }
    \G \F; (\G \initially
    \cnccsched);(\one \lor \X);\initially \neg \abscsched \Wide\land \G \F;\initially \cncfsched
    \hint{\limp}{ $\G \initially \abscsched$ }
    \G \F; (\G \initially
    \cnccsched);(\one \lor \X);(\initially \neg \abscsched \land
    \initially \abscsched) \Wide\land \G \F;\initially \cncfsched
    \hint{\limp}{ contradiction }
    \G \F; (\G \initially
    \cnccsched);(\one \lor \X);\cfalse \Wide\land \G \F;\initially \cncfsched
    \hint{\limp}{ $(\one \lor \X);\cfalse = \cfalse$ }
    \G \F; (\G \initially
    \cnccsched);\cfalse \Wide\land \G \F;\initially \cncfsched
    \hint{\limp}{ property of eternal computation \eqref{eq:eternal-false} }
    \G \F; (\G \initially
    \cnccsched) \land \E \Wide\land \G \F;\initially \cncfsched
    \hint{\limp}{ weakening }
    \G \F; (\G \initially
    \cnccsched) \Wide\land \G \F;\initially \cncfsched
    \hint{\limp}{ $\G$ is strengthening \eqref{eq:g-strengthen} }
    \F; (\G \initially
    \cnccsched) \Wide\land \G \F;\initially \cncfsched
    \hint{\limp}{ $\G \F;\initially \cncfsched$ is persistent
      \eqref{eq:g-persistent} and
      persistence rule \eqref{eq:persistence} }
    \F; (\G \initially \cnccsched \Wide\land \G \F;\initially \cncfsched)
  \end{calculation}
%

  Finally, the proof of \eqref{ref:sched} is as follows. Expanding the
  definition of $\schedule$, we prove
  \begin{align}
    \G (\G \initially \abscsched \land \G \F;\initially \absfsched  \Wide\limp
    \F;\absfsched;\action.(\absevt.\idx))\label{eq:3}
  \end{align}
  under the assumptions
  \begin{align}
    \G (\G \initially \cnccsched \land \G \F;\initially
    \cncfsched  &\Wide\limp
    \F;\cncfsched;\action.(\cncevt.\idx)) \label{eq:4}
  \end{align}

  First, notice that we drop the outer $\G$ from \eqref{eq:3},
  \eqref{eq:4}. and start the proof with the left-hand side of
  \eqref{eq:3}.
  \begin{calculation}
    \G \initially \abscsched \land \G \F;\initially \absfsched 
    \hint{\limp}{ \eqref{eq:1} }
    \F; (\G \initially \cnccsched \land \G \F;\initially \cncfsched))
    \hint{\limp}{ \eqref{eq:4} }
    \F;(\G \initially \cnccsched \land \F;\cncfsched;\action.(\cncevt.\idx))
    \hint{\limp}{ invariant property \eqref{eq:invariant-middle}, and
      \F;\F = \F }    
    \F;(\cnccsched \land \cncfsched);\action.(\cncevt.\idx)
    \hint{\limp}{ \eqref{ref:f-str} }
    \F;\absfsched;\action.(\cncevt.\idx)
    \hint{\limp}{ \eqref{ref:safety} }
    \F;\absfsched;\action.(\absevt.\idx)
  \end{calculation}
\end{proof}

\Thm~\ref{thm:schedule-ref} leads to the following proof rule.
\begin{equation}
  \label{rule:schedule-ref}
  \proofrule{
    \cncMch \WIDE\vdash \abscsched \land \absfsched \wide\leadsto
    \cnccsched \\
    \cncMch \WIDE\vdash \cnccsched \wide\un \spneg \abscsched \\
    \cncMch \WIDE\vdash \abscsched \land \absfsched \wide\leadsto
    \cncfsched \\
    \WIDE\vdash I_c \land \cnccsched \land \cncfsched
    \wide\limp \absfsched
  }
  {
    \cncMch \WIDE\vdash \schedule.(\cncevt.\idx) \wide\limp \schedule.(\absevt.\idx)
  }
\end{equation}

The following corollaries are direct consequences of
\Thm~\ref{thm:schedule-ref}, hence concerning events \absevt and
\cncevt as in \eqref{eq:absevt1} and \eqref{eq:cncevt1}.  They illustrate different ways of
refining event scheduling information: \emph{weakening the
  coarse schedule}, \emph{replacing the coarse schedule},
\emph{strengthening the fine schedule}, and \emph{removing the
  fine schedule}.

\begin{Corollary}[Coarse schedule weakening]
  Given $\cncfsched = \absfsched$, we have \[\schedule.(\cncevt.t) \limp \schedule.(\evt.t)\] if
  \begin{align}
    \G \initially(\abscsched \limp
      \cnccsched)~.\label{eq:c-wkn}
  \end{align}
\end{Corollary}
\begin{proof}[Sketch]
  Given $\cncfsched = \absfsched$, conditions \eqref{ref:f-flw} and
  \eqref{ref:f-str} of \Thm~\ref{thm:schedule-ref} are
  trivial. Conditions \eqref{ref:c-flw} and \eqref{ref:c-stb} are
  direct consequences of \eqref{eq:c-wkn}.
  
\end{proof}

\begin{Corollary}[Coarse schedule replacement]
  \label{cor:ref-rep-crs}
  Given $\cncfsched = \absfsched$, we have \[\schedule.(\cncevt.t) \limp \schedule.(\evt.t)\] if
  \begin{align}
    \abscsched \land \absfsched \wide\leadsto \cnccsched\tag{C\_FLW}\label{eq:csr-c-flw} \\
    \tag{C\_STB}\label{eq:csr-c-stb}
    \cnccsched ~\un~ \spneg \abscsched~.
  \end{align}
\end{Corollary}
\begin{proof}[Sketch]
  Given $\cncfsched = \absfsched$, conditions \eqref{ref:f-flw} and
  \eqref{ref:f-str} of \Thm~\ref{thm:schedule-ref} are trivial.
\end{proof}

\begin{Corollary}[Fine schedule strengthening]
  \label{cor:strengthen-fns}
  Given $\cnccsched = \abscsched$, we have \[\schedule.(\cncevt.\idx) \limp \schedule.(\absevt.\idx)\] if
  \begin{gather}
    \label{eq:fss-f-flw}
    \tag{F\_FLW}
    \abscsched \land \absfsched \wide\leadsto \cncfsched~\textrm{, and}\\
    \label{eq:fss-f-str}
    \tag{F\_STR}
    \G \initially(\cncfsched \limp\absfsched)~.
  \end{gather}
\end{Corollary}
\begin{proof}[Sketch] Given $\cnccsched = \abscsched$, conditions
  \eqref{ref:c-flw} and \eqref{ref:c-stb} of
  \Thm~\ref{thm:schedule-ref} are trivial.
\end{proof}

\begin{Corollary}[Fine schedule removal]
  \label{cor:remove-fns}
  Given $\cnccsched = \abscsched$ and $\cncfsched = \one$, we have
  \[\schedule.(\cncevt.\idx) \limp \schedule.(\absevt.\idx)\] if
  \begin{equation}
    \label{eq:remove-fns}
    \G \initially (\abscsched \limp \absfsched)~.
  \end{equation}
\end{Corollary}
\begin{proof}[Sketch]
  Given $\cnccsched = \abscsched$, conditions \eqref{ref:c-flw} and
  \eqref{ref:c-stb} of \Thm~\ref{thm:schedule-ref} are trivial.  Given $\cncfsched = \one$,
  condition \eqref{ref:f-flw} is trivial and condition
  \eqref{ref:f-str} is a direct consequent of \eqref{eq:remove-fns}.
\end{proof}

A special case of event refinement allows to remove event indices as
illustrated by the following theorem.
\begin{Theorem}[Events' indices removal] Consider \absevt as follows
  \label{thm:idx-rmv}  
  \begin{equation}
    \ubeventinlineidx{\absevt}{\idx,\cncidx}
    {\idx = E.\cncidx.\var \land \csched.(\idx,\cncidx).\var}
    {\fsched.(\idx,\cncidx).\var}
    {}{\guard.(\idx,\cncidx).\var}{\assignment.(\idx,\cncidx).\var.\var'}
    \label{eq:absevt2}
  \end{equation}
  The following event \cncevt is a refinement of \absevt, i.e.,
  satisfying \eqref{eq:evt-safety} and \eqref{eq:evt-live}.
  \begin{equation}
    \ubeventinlineidx{\cncevt}{\cncidx}
    {\csched.(E.\cncidx.\var,~\cncidx).\var}
    {\fsched.(E.\cncidx.\var,~\cncidx).\var}
    {}{\guard.(E.\cncidx.\var,~\cncidx).\var}
    {\assignment.(E.\cncidx.\var,~\cncidx).\var.\var'}
    \label{eq:cncevt2}
  \end{equation}
\end{Theorem}
\begin{proof}
  For \eqref{eq:evt-safety}, we use $E.\cncidx.\var$ as the witness
  for the removing indices $\idx$, which leads to the proof
   obligation: \[\action(\cncevt.j) \wide\limp
  \action(\absevt.(E.\cncidx.\var,j))~.\] For \eqref{eq:evt-live}, we
  note that in the case where $\idx \neq E.\cncidx.\var$, the coarse
  schedule of \absevt is $\cfalse$, hence \absevt.\idx is unscheduled,
  hence \eqref{eq:evt-live} is satisfied. Therefore,
  $\schedule.(\cncevt.\cncidx) =
  \schedule(\absevt.(E.\cncidx.\var,~\cncidx))$ holds.
\end{proof}


\newcommand{\PREAMBLE}

\section{Example: A Signal Control System}
\label{sec:example}

We illustrate our method by applying it to design a system controlling
trains at a
station~\cite{hudon11:_devel_contr_system_guided_model_envir}.  We
first present some informal requirements of the system.

\subsection{Requirements}
\label{sec:requirements}

The network at the station contains an \emph{entry block}, several
\emph{platform blocks} and an \emph{exiting block}, as seen in
Figure~\ref{fig:sgnctrl}.  Trains arrive on the network at the entry
block, then can move into one of the platform blocks before moving to
the exiting block and leaving the network.
In order to control the trains at the station, signals are positioned
at the end of the entry block and each platform block.  The
train drivers are assumed to obey the signals.  The signals are
supposed to change from green to red automatically when a train passes by.
\begin{figure*}[!htbp]
  \centering%
  \ifx\PREAMBLE\UnDef
\documentclass{beamer}
\usepackage{tikz}
\usetikzlibrary{snakes,arrows}

\usepackage[english]{babel}

\usepackage[latin1]{inputenc}

\usepackage[T1]{fontenc}
\usepackage{amssymb}
\usepackage{amsmath}
\usepackage{eventB}

\begin{document}
\else
\fi

\begin{tikzpicture}[scale=1]
  \draw[very thick] (0,0) -- (1.5,0);
  \draw (0.7, -1) node{entry block};

  \draw[very thick] (1.5,0) -- (2,0);
  \draw[very thick, dashed] (2,0) -- (2.5,1.5);
  \draw[very thick, dashed] (2,0) -- (2.5,-1.5);
  \draw[very thick, dashed] (2,0) -- (2.5,0);
  
  \draw[very thick] (2.6,0) -- (4.9,0);
  \draw[very thick] (2.6,1.5) -- (4.9,1.5);
  \draw[very thick] (2.6,-1.5) -- (4.9,-1.5);
  \draw (3.75, -2.5) node{platform blocks};

  \draw[very thick] (5.5,0) -- (6,0);
  \draw[very thick, dashed] (5,1.5) -- (5.5,0);
  \draw[very thick, dashed] (5,0) -- (5.5,0);
  \draw[ very thick, dashed] (5,-1.5) -- (5.5,0);

  \draw[very thick] (6.0,0) -- (7.5,0);
  \draw (6.8, -1) node{exit block};

  \draw (1, 0.2) rectangle +(0.3,0.7);
  \filldraw[red] (1.15, 0.7) circle (0.1);
  \filldraw[green] (1.15, 0.4) circle (0.1);
  \draw (1.15, 1.2) node{entry signal};

  \draw (4.4, 0.2) rectangle +(0.3,0.7);
  \filldraw[red] (4.55, 0.7) circle (0.1);
  \filldraw[green] (4.55, 0.4) circle (0.1);
  \draw (4.55, 2.9) node{platform signals};

  \draw (4.4, 1.7) rectangle +(0.3,0.7);
  \filldraw[red] (4.55, 2.2) circle (0.1);
  \filldraw[green] (4.55, 1.9) circle (0.1);

  \draw (4.4, -1.3) rectangle +(0.3,0.7);
  \filldraw[red] (4.55, -0.8) circle (0.1);
  \filldraw[green] (4.55, -1.1) circle (0.1);

  \draw[dotted] (-2,0) -- (0, 0);
  \draw(-1.5,0.2) rectangle +(1.1, 0.3);
  \draw(-1.5,0.5) rectangle +(0.3, 0.3);
  \draw(-0.8,0.5) rectangle +(0.1, 0.2);
  \draw[decorate, decoration={snake, amplitude = 1pt, segment length = 2pt}] (-0.8,0.7) --  (-1,1);
  \draw[decorate, decoration={snake, amplitude = 1pt, segment length = 2pt}] (-0.75,0.7) --  (-0.95,1);
  \draw[decorate, decoration={snake, amplitude = 1pt, segment length = 2pt}] (-0.7,0.7) --  (-0.9,1);
  \draw (-1.4, 0.1) circle (0.1);
  \draw (-1.1, 0.1) circle (0.1);
  \draw (-0.8, 0.1) circle (0.1);
  \draw (-0.5, 0.1) circle (0.1);
  \draw (-1, 1.2) node{$\Longrightarrow$};
\end{tikzpicture}

\ifx\PREAMBLE\UnDef
\end{document}
\else
\fi
  \caption{A signal control system}
  \label{fig:sgnctrl}
\end{figure*}
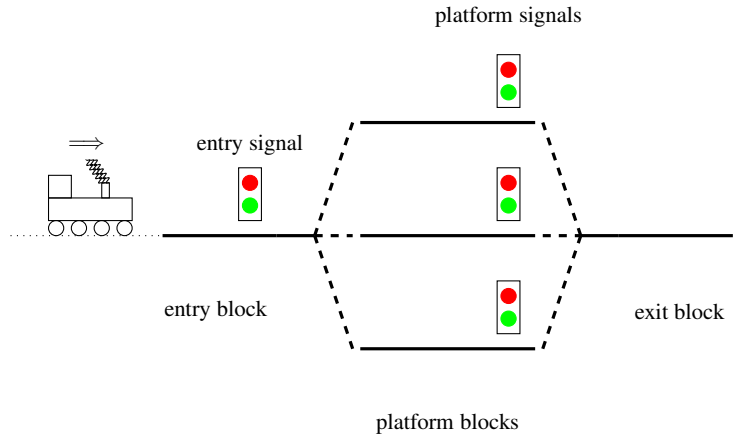

The most important properties of the system are that (1) there should be
no collision between trains (\ref{saf:no-collision}), and (2) each train in the network eventually
leaves~(\ref{live:train-leave}).
\begin{requirements}
  \saf{saf:no-collision}{There is at most one train on each block}\ReqSpacing
  \fun{live:train-leave}{Each train in the network eventually leaves}\ReqSpacing
  \env{env:top}{The tracks are arranged according to Figure~\ref{fig:sgnctrl}}
  \fun{saf:top}{Every train enters only through the \emph{entry block}, 
  	then proceed to a \emph{platform block} and move on to the 
	\emph{exit block} from where they leave the station. }
  \eqp{eqp:sgn}{ A light signal is positioned after the entrance block
  	and after each of the platforms. }
  \env{env:obey}{ Train drivers obey the light signals, i.e. when the
  	signal is green, they advance and they stop when the 
	signal is red. }
\end{requirements}

\paragraph{Refinement strategy} 
Our development consists of an initial model and five refinement
steps.  We summarize our refinement strategy for developing the signal
control system as follows.
\begin{description}
\item[Init. model] We abstractly model the trains in the network,
  focusing on~\ref{live:train-leave}.
\item[1st Ref.]  We introduce the topology of the
  network~\ref{env:top} and \ref{saf:top}.
\item[2nd Ref.]  We strengthen the model of the system, focusing
  on~\ref{saf:no-collision}.
\item[3rd Ref.] We introduce the signals and derive a specification
  for the controller that manages these signals~\ref{eqp:sgn} and
  \ref{env:obey}.
\item[4th Ref.] We refine the controller's specification, in
  particular, scheduling the trains passing the station in a
  first-in-first-out manner.
\item[5th Ref.] We refine further the controller's specification so
  that it can be implemented in some programming language.
\end{description}

\newcommand{\D}{{\cal{D}}}
\paragraph{Notation} Well-definedness \cite{Mehta2008} is an important issue when 
	dealing with partial functions. However, when trying to make
	formulas well-defined, some overhead often has to be 
	introduced which can make said formulas bulky. 
	For example, if
	we need to express 
	\[ f.x \le g.y~,\] 
	with $f,g$ two partial functions,
	the formula is only meaningful in the case where $x \in \dom.f 
	\land y \in \dom.g$ and the formula above is therefore not 
	necessarily well defined. This new formula
	\[ x \in \dom.f \1\land y \in \dom.g \1\land f.x \le g.y \] 
	is well-defined and is often a suitable substitute for $f.x \le g.y$ 
	but it is much longer and the subject matter, the ordering of $f$ 
	and $g$, constitute only a small fraction of the formula: the 
	attention of the reader is mostly drawn to the technicality of
	well-definedness.
	
	As a shorthand, we will use a new notation defined as $\langle 
	P \rangle \triangleq \D(P) \land P$. In the previous example, we 
	can express the property as 
	\[ \langle f.x \le g.y \rangle \] 
	which is false if $f.x$ or $g.y$ is ill-defined and has
	the expected truth value otherwise. It might be also handy to 
	have a shorthand for $x \in \dom.f \land y \in \dom.g \1\implies 
	f.x \le g.y$ ---
	which is true if $f.x$ or $g.y$ is ill-defined and has the normal 
	truth value otherwise --- but we won't need it in this paper 
	and therefore refrain from defining a shorthand for it.

\paragraph{Logic}
In Sect.~\ref{sec:contribution}, we conducted the proofs of soundness of 
the refinement rules and the inference rules of temporal properties using
computation calculus. In the example, we will conduct our reasoning 
using these inference rules and predicate calculus without reference to
computation calculus. The purpose is to use the rules as a clear interface
between the semantics of \unitb and the reasoning about \unitb models.

\paragraph{Proof Format}
The equational proof format has been used to advantage already in
Section~\ref{sec:background} and Section~\ref{sec:contribution}.  There we
use this format in rewriting an expression with value preserving rules
or various order preserving rules (e.g. $\implies$, $\le$), from an
initial expression to a final expression.  The style of manipulations
has an algebraic flavour.  In this section, we use equational proof
format to manipulate sequents instead of normal expressions. While an
expression has a value, a sequent is provable or not. The usual way of
relating two sequents is the inference rule:
\begin{align*}
\proofrule{ \Gamma, \alpha \vdash \phi  }{ \Gamma \vdash \psi }
\end{align*}
When building a formal proof with them, the format becomes quickly
unwieldy and unreadable. Instead, we use $\sqsubseteq$ to relate two
sequents in equational proofs with the understanding that
$ \Gamma \vdash \psi \1\sqsubseteq \Gamma, \alpha \vdash \phi $ 
stands for the inference above, i.e. $\Gamma \vdash \psi$ has a proof
if $\Gamma, \alpha \vdash \phi$ has a proof.

Sometimes, inferences rules have more than one premise. In such cases,
in our calculations, either we keep the most important one as the main thread 
of reasoning and refer to the other ones in the hint of the step, or we list
the ones we kept one above the other. The subsequent steps can apply to
any one of them.

\paragraph{Naming Convention}
We adopt the following convention in naming the properties appearing
in the subsequent development to indicate the type of the property
(e.g., invariance, unless, or progress), the level of refinement, and
the sequent number of the property.  For example, $\Binv{inv0\_1}$ is
the 1st invariant of the initial model, while $\Binv{un2\_1}$ and
$\Binv{prg2\_1}$ are the 1st unless property and the 1st progress
property of the 2nd refinement, respectively.

\paragraph{Refinement Strategy}
The refinement strategy for our development is as follows.
\begin{description}
\item[Initial Model] focuses on specifying and reasoning about the
  main progress requirement~\ref{live:train-leave}.
\item[First refinement] introduces the topology of the train
  station~\ref{env:top} and the movement of the trains through the
  station \ref{saf:top}.
\item[Second Refinement] incorporates the safety requirements of the
  system to prevent train collisions~\ref{saf:no-collision}.
\item[Third Refinement] adds the light signals to the
  model~\ref{eqp:sgn} and the assumption that the train drivers always
  obey these light signals~\ref{env:obey}.
\item[Fourth Refinement] realises a software controller for the
  signals in the station.  In particular, the scheduling of the train
  passing through the station is performed using a queue.
\item[Fifth Refinement] simplifies the software controller by removing
  the index of the corresponding event.
\end{description}

\subsection{Initial Model \InitM --- Arriving and Departing}
\label{sec:initial-model}
In this initial model \InitM, we use a carrier set \TRAIN to denote
the set of trains and a variable \trains (short for station) to denote
the set of trains currently inside the station.
\begin{Bcode}
  $\variables{\trains}$
  \Bhspace
  $
  \invariants{\Binv{inv0\_1}: & \trains \subseteq \TRAIN}
  $
\end{Bcode}
Initially \trains is assigned the empty set $\emptyset$.  At this
abstract level, we have two events to model a train arriving at the
station and a train leaving the station as follows:
\begin{Bcode}
  $
  \ubeventinlineidx{\arrive}
  {\train}{}{}{}
  {\train \in \TRAIN}
  {\trains \bcmeq \trains \bunion \{\train\}}
  $
  \Bhspace
  $
  \ubeventinlineidx{\depart}{\train}{}{}{}{\train \in \TRAIN}{\trains \bcmeq
    \trains \setminus \{\train\}}
  $
\end{Bcode}

Requirement \ref{live:train-leave} can be specified as a progress
property (with $\train$ implicitly quantified universally over the whole
property): %
\begin{equation}
  \label{eq:prg0-1}
  \train \in \trains ~\leadsto~ \lnot t \in \trains~.
  \tag{\Binv{prg0\_1}}
\end{equation}

We attempt to use event \depart to implement \ref{eq:prg0-1} as follows.
\begin{calculation}
  \InitM \Wide\vdash \train \in \trains \leadsto \lnot \train \in
  \trains
  \hint{\sqsubseteq}{ Transient definition \eqref{eq:transient}}
  \InitM \Wide\vdash \tr~\train \in \trains
  \hint{\sqsubseteq}{ Transient rule~\eqref{eq:tr-rule} with \depart.\train}
  \InitM \Wide\vdash \depart.\train \2\falsifies \train \in \trains
  \hint{\sqsubseteq}{ Falsifies rule~\eqref{eq:falsifies} with: \eqref{eq:SCH1} and \eqref{eq:NEG1} }
  \true
\end{calculation}
with:
\begin{align*}
  \label{eq:SCH1}
  \tag{C\_EN\_1} & t \in \trains \1\implies \cfalse \\
  \label{eq:NEG1}
  \tag{NEG\_1} & t \in \trains \land \cfalse \land \trains' \!=\! \trains \setminus \{ t \} 
  \1\implies \neg t \in \trains' 
\end{align*}

\begin{sloppypar}
The proof obligation \eqref{eq:NEG1} is trivial.  However, \eqref{eq:SCH1}
cannot be proved because the coarse schedule of \depart is $\cfalse$ (since
\depart is current unscheduled).  We can remedy this situation by
adding a coarse schedule to \depart, which becomes as follows:
\end{sloppypar}
\begin{Bcode}
  $
  \ubeventinlineidx{\depart}{\train}{\train \in \trains}{}{}{\train \in \TRAIN}{\trains \bcmeq \trains \setminus \{\train\}}
  $
\end{Bcode}
The updated proof obligations are:
\begin{align*}
  \label{eq:SCH1:prime}
  \tag{C\_EN\_1'} & \train \in \trains \1\implies \train \in \trains \\
  \label{eq:NEG:prime}
  \tag{NEG\_1'} & \train \in \trains 
  \1\land \train \in \TRAIN 
  \1\land \trains' \!=\! \trains \setminus \{ \train \} 
  \2\implies \neg \train \in \trains'~. 
\end{align*}
The proof obligations \eqref{eq:SCH1:prime} and \eqref{eq:NEG:prime}
can be easily discharged.

Note that event \depart has different guard and coarse schedule.  It
is our intention to design \depart with a weak guard and a strong
coarse schedule that allow us to prove system properties (e.g.,
invariance and progress properties).  This gives more flexibility in
strengthening events' guards and weakening schedules as needed during
the course of refinement.%

Since event \arrive will not affect the reasoning about progress
properties (it is always unscheduled), we are going to omit its
refinement in the subsequent presentation.

\subsection{First Refinement \MchI --- The Topology} 
\label{sec:first-refinement}
In this refinement \MchI, we first introduce the topology of the network
in terms of blocks (\ref{env:top}).  We introduce a carrier set $\BLOCK = 
 \{\Entry\} \bunion \PLATFORM \bunion \{ \Exit \}$ denoting 
the entry block, the platform blocks and the exit block, respectively.
A new variable \location is added to denote the location of trains
in the network, constrained by this invariant:
\begin{gather} 
\label{eq:inv1-1}
\tag{\Binv{\invLocationType}} \location\in \trains \tfun \BLOCK.
\end{gather}

To capture \ref{saf:top}, we formulate the following safety properties:
\begin{align}
	\label{eq:saf1-1} \tag{\Binv{un1\_1}}
	\neg \train \in \trains \2{&\un} \langle \location.\train = \Entry \rangle \\
	\label{eq:saf1-2} \tag{\Binv{un1\_2}}
	\langle \location.\train = \Entry \rangle 
		\2{&\un} \langle \location.\train \in \PLATFORM \rangle \\
	\label{eq:saf1-3} \tag{\Binv{un1\_3}}
	\langle \location.\train \in \PLATFORM \rangle 
		\2{&\un} \langle \location.\train = \Exit \rangle \\
	\label{eq:saf1-4} \tag{\Binv{un1\_4}}
	\langle \location.\train = \Exit \rangle 
		\2{&\un} \neg \train \in \trains
\end{align}
They can be summarized in Figure~\ref{fig:transition}.
\begin{figure}[!htbp]
  \centering
  \ifx\PREAMBLE\UnDef
\documentclass{beamer}
\usepackage{tikz}
\usetikzlibrary{shapes}
\usepackage[english]{babel}

\usepackage[latin1]{inputenc}
\usepackage[color]{eventB}
\usepackage{sgnctrl}

\begin{document}
\else
\fi
\begin{tikzpicture}[scale=0.5]
  \scriptsize
  \draw (0,0) node(entry)[circle, draw, minimum height = 4ex, minimum width=7em]{$\location.\zug = \Entry$};
  \draw (5,0) node(plf)[circle, draw, minimum height = 4ex, minimum
  width=7em]{$\location.\zug \in \PLATFORM$};
  \draw (10,0) node(exit)[circle, draw, minimum height = 4ex, minimum width=7em]{$\location.\zug = \Exit$};
  \draw[dashed] (-3,-2.5) rectangle +(16,5.5);
  \draw (-2,-2) node{$\zug \in \trains$};
  \draw (12,-2) node{$\zug \in \trains$};

  \draw (5,-4.5) node(out)[circle, draw, minimum height = 4ex, minimum width=7em]{$\zug \notin \trains$};


  \draw[->, very thick] (out) -- (entry);
  \draw[->, very thick] (entry) .. controls (-1,3) and (1,3) .. (entry);
  \draw[->, very thick] (entry) .. controls (2,2) and (3,2) .. (plf);
  \draw[->, very thick] (plf) .. controls (4,3) and (6,3) .. (plf);
  \draw[->, very thick] (plf) .. controls (7,2) and (8,2) .. (exit);
  \draw[->, very thick] (exit) .. controls (9,3) and (11,3) .. (exit);
  \draw[->, very thick] (exit) -- (out);
  \draw[->, very thick] (out) .. controls (6,-7.5) and (4,-7.5) .. (out);
\end{tikzpicture}

\ifx\PREAMBLE\UnDef
\end{document}
\else
\fi
  \caption{State transitions for trains}
  \label{fig:transition}
\end{figure}

We use \eqref{eq:un-rule} to prove that they hold.  In particular, for
\ref{eq:saf1-2} and \ref{eq:saf1-3}, we need to strengthen the guard of \depart.
Subsequently, in order to make sure that the schedule is stronger than
the guard~(condition \eqref{eq:fis}), we need to strengthen the
coarse schedule accordingly.  An assignment for \location is added for
the maintenance of \ref{eq:inv1-1}.
\begin{Bcode}
  $%
  \ubeventidx{\depart}{\train}{}{%
    \train \in \trains \1\land \location.\train = \Exit%
  }%
  {\train \in \trains \1\land \location.\train = \Exit}%
  {}%
  {%
    \trains\bcmeq \trains \setminus \{\train\} \\
    \location \bcmeq \{\train\} \domsub \location%
  }%
  $
\end{Bcode}

In order to prove the refinement of \depart, we apply
\Cor~\ref{cor:ref-rep-crs} (coarse schedule replacing).  In
particular, conditions \eqref{eq:csr-c-flw} and \eqref{eq:csr-c-stb}
require us to prove the following properties:
\begin{gather}
  \train \in \trains ~\leadsto~ 
  \langle \location.\train = \Exit \rangle \label{eq:prg1-5}
  \tag{\Binv{prg1\_1}} \\
  \langle \location.\train =
     \Exit \rangle ~\un~ \neg \train \in \trains \label{eq:saf1-4}
  \tag{\Binv{un1\_4}}
\end{gather}

From now on, we focus on reasoning about progress properties, e.g.,
\ref{eq:prg1-5}, omitting the reasoning about unless properties, e.g.,
\ref{eq:saf1-4}.  The proofs of these unless properties can be done using
\eqref{eq:un-rule} and will be omitted here.

In order to satisfy \ref{eq:prg1-5}, we first transform it into a
transient property.
\begin{calculation}
		\train \in \trains ~\leadsto~ 
  		\langle \location.\train = \Exit \rangle
	\hint{=}{ \eqref{eq:37} }
		\langle \neg~ \location.\train = \Exit \rangle ~\leadsto~ 
  		\langle \location.\train = \Exit \rangle
	\hint{\sqsubseteq}{ \eqref{eq:transitivity} }
		\langle \neg~ \location.\train = \Exit \rangle ~\leadsto~ 
  		\langle \location.\train \in \PLATFORM \rangle 
\\ &		\langle \location.\train \in \PLATFORM \rangle ~\leadsto~ 
  		\langle \location.\train = \Exit \rangle
                \mhint[.65\columnwidth]{\sqsubseteq}{ \eqref{eq:37} and with \eqref{eq:prg1-6} 
                  \\ (see below) on first leads-to property }
		\langle \location.\train \in \PLATFORM \rangle ~\leadsto~ 
  		\langle \location.\train = \Exit \rangle 
                \hint{\sqsubseteq}{ \eqref{eq:ensure} with \eqref{eq:saf1-3} }
		\tr ~ \langle \location.\train \in \PLATFORM \rangle
\end{calculation}
with the new property:
\begin{gather}
\label{eq:prg1-6} \tag{\Binv{prg1\_2}}
	\langle \location.\train = \Entry \rangle ~\leadsto~ 
	\langle \location.\train \in \PLATFORM \rangle
\end{gather}

\begin{sloppypar}
We implement the resulting transient property, i.e., $\tr ~ \langle
\location.\train \in \PLATFORM \rangle$ using a new event
\moveout.  We leave the coarse and the fine schedule as some unknown
$\csched?$ and $\fsched?$, and see how to derive them from resulting
proof obligations.
\end{sloppypar}
\begin{Bcode}
  $%
  \ubeventinlineidx{\moveout}{\train}
  {\csched?}%
  {\fsched?}%
  {}{%
    \begin{array}[t]{l}
      \train \in \trains \wide\land\\
      \location.\train \in \PLATFORM%
    \end{array}
  }%
  {\location.\train \bcmeq \Exit}%
  $
\end{Bcode}
The proof that \moveout implements the transient property is as
follows.
\begin{calculation}
  \MchI \Wide\vdash \tr ~ \langle \location.\train \in \PLATFORM \rangle
  \hint{\sqsubseteq}{ Transient rule~\eqref{eq:tr-rule} with \moveout.\train}
  \MchI \Wide\vdash \moveout.\train \2\falsifies \langle \location.\train \in \PLATFORM \rangle
  \hint{=}{ Falsifies rule~\eqref{eq:falsifies} with \eqref{eq:sch2?} and \eqref{eq:neg2?} }
  \MchI \Wide\vdash \langle \location.\train \in \PLATFORM \rangle
  \land \csched? \wide\leadsto \fsched?
\end{calculation}
where:
\begin{gather}
  \tag{C\_EN\_2}\label{eq:sch2?}%
  \begin{array}{ll}
    \phantom{\implies}\quad & \langle \location.\train \in \PLATFORM \rangle \2\implies \csched?
  \end{array}
  \\[2ex]%
  \tag{NEG\_2}\label{eq:neg2?}%
  \begin{array}{ll}
    & \langle \location.\train \in \PLATFORM \rangle \wide\land  \csched? \wide\land
    \fsched?  \wide\land \\
    & \location' \2= \location \ovl \{ \train \mapsto \Exit \}\footnotemark{} \wide\land \trains' = \trains \\
    \implies \quad & \neg \langle \location'.\train \in \PLATFORM \rangle
  \end{array}
\end{gather}
\footnotetext{%
  $\location \ovl \{\train \mapsto \Exit \}$ denotes a relation equal
  to \location excepts for the entry for \train which is mapped to \Exit.}

We design the coarse schedule $\csched?$ and the fine schedule
$\fsched?$ such that the goal, i.e., $\langle \location.\train
\in \PLATFORM \rangle \land \csched? \wide\leadsto \fsched?$, and
conditions \eqref{eq:sch2?} and \eqref{eq:neg2?}  can be discharged
trivially.  One such design is to have $\csched?$ being $\train \in
\trains \1\land \location.\train \in \PLATFORM$ and $\fsched?$ being
$\true$.  This gives us the following design for \moveout:
\begin{Bcode}
  $%
  \ubeventidx{\moveout}{\train}{}{%
    \train \in \trains \land \location.\train \in \PLATFORM%
  }%
  {\train \in \trains \land \location.\train \in \PLATFORM}%
  {}%
  {\location.\train \bcmeq \Exit}%
  $
\end{Bcode}

The updated conditions \eqref{eq:sch2?} and \eqref{eq:neg2?} is as
follows and can be discharged easily.
\begin{gather}
  \tag{C\_EN\_2}\label{eq:sch2}
  \begin{array}{ll}
    \phantom{\implies} \quad & \langle \location.\train \in \PLATFORM \rangle 
    \wide\implies \train \in \trains \1\land \location.\train \in \PLATFORM  
  \end{array}
  \\[2ex] \tag{NEG\_2}\label{eq:neg2}
  \begin{array}{ll}
    & \langle \location.\train \in \PLATFORM \rangle 
    \wide\land \\
    & \train \in \trains \1\land \location.\train \in \PLATFORM 
    \wide\land \true \wide\land\\
     & \location' \2= \location \ovl \{ \train \mapsto \Exit \}
     \wide\land \trains' = \trains \\
    \implies \quad & \neg \langle \location'.\train \in \PLATFORM \rangle
  \end{array}
\end{gather}

The remaining progress property, i.e., \ref{eq:prg1-6}, can be
implemented in a similar fashion.  We first transform \ref{eq:prg1-6}
into a transient property and implement it by the following new event
\movein.
\begin{Bcode}
  $%
  \ubeventidx{\movein}{\train}{}%
  {\train \in \trains \land \location.\train = \Entry}%
  {\train \in \trains \land \location.\train = \Entry}%
  {}%
  {\location.\train \bcmin \PLATFORM }
  $
\end{Bcode}

Applying the unless rule \eqref{eq:un-rule}, we can verify that the
new events, i.e., \movein and \moveout, satisfy safety constraints,
such as \ref{eq:saf1-1}, \ref{eq:saf1-2}, \ref{eq:saf1-3}, and
\ref{eq:saf1-4}.  It should be noted that those safety requirements
guided our design of the new events along side the progress properties
that they are meant to satisfy.
%
%
%
\subsection{Second Refinement \MchII --- Preventing Collisions}
\label{sec:second-refinement}
In this refinement, \MchII, we incorporate the safety requirement stating that
there are no collisions between trains within the network,
i.e., \ref{saf:no-collision}.  This is captured by a new invariant 
about \location:
\begin{gather}
\tag{\Binv{inv2\_1}} \qforall{\train_1,\train_2}{\train_1, \train_2 \in \trains \land \location.\train_1 =
  \location.\train_2}{\train_1 = \train_2}.
\end{gather}

\begin{sloppypar}
The guard of event \moveout needs to be strengthened with the fact
that the exit block is free (i.e., $\neg \Exit \in \ran.\location$),
to maintain \Binv{inv2\_1}.  Due to the feasibility
condition~\eqref{eq:fis} for \unitb events (requiring the schedules
to be stronger than the guard), we need to strengthen the schedules
accordingly.  
In particular, we add a fine schedule to \moveout:
\end{sloppypar}
\begin{Bcode}
  $%
  \ubeventidx{\moveout}{\train}{}%
  {%
    \train \in \trains \wide\land
    \location.\train \in \PLATFORM \wide\land
    \neg \Exit \in \ran.\location%
  }%
  {\train \in \trains \land \location.\train \in \PLATFORM}%
  {\neg \Exit \in \ran.\location} {\location.\train \bcmeq \Exit}%
  $
\end{Bcode}
The scheduling information for \moveout states that for any train
$\train$, if $\train$ stays in a platform for infinitely long and the
exit block becomes free infinitely often, then $\train$ will
eventually move out of the platform. 

\paragraph{Heuristics.}
So far, all the scheduling information was encoded in coarse schedules.
It is a general principle that coarse schedules should be used over
fine schedules whenever possible. This is because, contrary to fine schedules, 
each coarse schedule can be manipulated in separation from each other. 

For example, if during refinement we want to replace the coarse schedule 
$a_0 \land a_1 \land p$ with $c_0 \land c_1 \land p$, we can meet the 
proof obligation by proving
\begin{align*}
a_0 \land a_1 \land p \leadsto c_0 \land c_1 \land p~,
\\ c_0 \land c_1 \land p \1\un a_0 \land a_1 \land p~. 
\end{align*}
Since the two schedules involved can be arbitrarily complex, it is often more 
convenient to break down the proof obligation into the following smaller ones:
\begin{align*}
    a_0 \leadsto& c_0
\\  a_1 \leadsto& c_1
\\  c_0 \1\un& \neg a_0
\\  c_1 \1\un& \neg a_1
\end{align*}
Instead of seeing the replacement of the coarse schedule as one 
operation, we can see it as the replacement of $a_0$ with $c_0$ 
and $a_1$ with $c_1$.
This is especially convenient because 
the set of concrete schedules is rarely manipulated all at once.

In comparison, when refining fine schedules, the proof obligation ($a_0 \land a_
1 \land p \leadsto c_0 \land c_1 \land p$) has to be dealt with as a whole. 
The comparison between coarse and fine schedule is similar when proving 
liveness properties. 
This is why we should keep the fine schedules as small as possible.

One situation favors fine schedules. Contentions happen when many events have 
to occur and the occurrence of one falsifies the schedules of the others. A 
mutual exclusion protocol is a good example of contention. The $
enter\_critical\_section$ event of each process has to occur when the process 
is waiting but no other process is in its critical section. When two processes, 
say $p_0$ and $p_1$ are waiting, if $p_0$ first enters its critical section, it 
falsifies $p_1$'s schedule. If it were a coarse schedule, this means 
that there is no guarantee that the other process will ever be granted access 
to its critical section. The situation can be fixed by making part of the 
events' schedule coarse and the other part fine. The fact that $p_1$ is waiting 
is stable and no other process will falsify this. It can therefore safely be made 
into the coarse schedule. The other part, the condition that no other process be
in their critical region, should be made into the fine schedule. It is not 
stable but as long as it becomes true infinitely often, $p_1$ will be granted 
access to its critical section.

As a rule of thumb, most schedules should be made into coarse schedules. 
When liveness properties cannot be proved, coarse schedules should be selectively 
made into fine schedules. 
\\ (end of \emph{heuristics})

In order to prove the refinement of
\moveout, we apply \Cor~\ref{cor:strengthen-fns} (fine schedule
strengthening), which requires to prove the following progress
property.  The abstract event \moveout has no fine schedules, it is
assumed to be $\true$.  Condition \eqref{eq:fss-f-str} is trivial
since the abstract fine schedule is $true$.  Condition
\eqref{eq:fss-f-flw} leads to the following property:
\begin{equation}
\langle \location.\train \in \PLATFORM \rangle \wide\leadsto \neg
\Exit \in \ran.\location~,\tag{\Binv{prg2\_1}}
\end{equation}
which can be strengthened to
\begin{equation}
  \label{eq:prg2-3}
  \true \wide\leadsto \neg \Exit \in \ran.\location
  \tag{\Binv{prg2\_2}}
\end{equation}
We satisfy \ref{eq:prg2-3} (which is a transient property) by applying
the transient rule \eqref{eq:tr-rule} using event \depart with the
index denoting the train at the $\Exit$ location, i.e.,
$\location^{-1}.\Exit$.  Intuitively, the train at the exit block will
eventually depart, hence the exit block becomes free.
\begin{calculation}
  \MchII \Wide\vdash \tr~ \Exit \in \ran.\location
  \hint{\sqsubseteq}{ Transient rule~\eqref{eq:tr-rule} with $\depart.((\location^{-1}).\Exit)$}
  \MchII \Wide\vdash \depart.((\location^{-1}).\Exit) \1\falsifies \Exit \in \ran.\location
  \hint{=}{ Falsifies rule~\eqref{eq:falsifies} with \eqref{eq:sch3}
    and \eqref{eq:neg3} }
  \true
\end{calculation}
where:
\begin{align*}
  \label{eq:sch3} \tag{C\_EN\_3}
  &\begin{array}{rll}
    &&\Exit \in \ran.\location \\
    \implies \\
    & & (\location^{-1}).\Exit \in \trains \\
    &\land & \location.(~(\location^{-1}).\Exit~) = \Exit 
  \end{array} \\[2ex]
  \label{eq:neg3} \tag{NEG\_3}
&\begin{array}{rll}
	& \Exit \in \ran.\location & \\
	\land & (\location^{-1}).\Exit \1\in \trains \\
	\land & \location.(~(\location^{-1}).\Exit~) = \Exit  \\
	\land & \true &  \\
	\land & \location' \1= \{ ~ (\location^{-1}).\Exit  ~\} \1
		\domsub \location \\
	\land & \trains' \1= \trains \1\setminus 
		\{ ~ (\location^{-1}).\Exit ~\} \\
\implies \quad & \neg ~ \Exit \in \ran.\location' &
\end{array}
\end{align*}
The proofs of conditions~\eqref{eq:sch3} and \eqref{eq:neg3} are
straightforward and will be left out.

Finally we strengthen the guard of \movein and subsequently strengthen
its coarse schedule.  We apply \Cor~\ref{cor:ref-rep-crs}
(coarse schedule replacing) \movein.  The detailed proof is omitted
here.
\begin{Bcode}
  $%
  \ubeventidx{\movein}{\train}{}%
  {%
    \train \in \trains \land \location.\train = \Entry 
    \land ~ \neg \PLATFORM \subseteq \ran.\location %
  }%
  {%
     \train \in \trains \land \location.\train = \Entry
    \land ~ \neg \PLATFORM \subseteq \ran.\location %
  }%
  {}%
  {\location.\train \bcmin \PLATFORM \setminus \ran.\location }%
  $
  \end{Bcode}

\subsubsection{Comparison between Coarse/Fine Schedules and
  Weak/Strong Fairness}
\label{sec:cf:sched:vs:w:str:fair}

Event \moveout has both a coarse and a fine schedule.  The alternative,
using only weak or strong fairness, would complicate the proofs
and make refinement of the system more difficult.

On the one hand, weak-fairness requires for the exit block to remain
free continuously in order for trains to move out.  This assumption is
not met by the current system: if, infinitely often, another train
than $\train$ located at a different platform moves on to the exit
block before $\train$ does, $\train$'s weak-fairness allows for
$\train$ to stay where it is forever.  In other words, the weak-
fairness assumption for \moveout will be \emph{too weak}; it does not
guarantee that a train inside the station will eventually exit.  An
attempt to prove the refinement with the weakly-fair \moveout event
using \Cor~{\ref{cor:ref-rep-crs}} will lead to the following
unprovable \eqref{eq:csr-c-stb} condition.
\begin{equation}
  \begin{array}{ll}
    & \langle \location.\train \in \PLATFORM \wide\land \neg
    \Exit \in \ran.\location \rangle \\
    \un\quad & \neg(\train \in \trains \wide\land \location.\train \in \PLATFORM)
  \end{array}\label{eq:c-stb-fail}
\end{equation}
The event that fails to satisfy \eqref{eq:c-stb-fail} is $\moveout$
for train other than the current $\train$.

On the other hand, strong-fairness would allow a train to access the
exit block if it is present on the platform intermittently.  This
assumption is more flexible than we need since it allows behaviours
where a train hops on and off the platform infinitely often while
waiting for its turn at the exit block.  The price of that flexibility
is to entangle properties of the exit block with properties of trains:
indeed, we would need not only to prove that the train will be on its
platform and that the exit block will become free but that both happen
simultaneously infinitely often.  More formally, while we can prove
that the strongly-fair \moveout event refines the abstract \moveout
event, future refinement of \moveout will be more difficult due to the
stronger scheduling assumption.  We choose to relinquish this
flexibility and are therefore capable of structuring our proof better:
on one hand, the train stays on its platform as long as necessary;
independently, the exit block becomes free infinitely many times.
This (choosing a weaker scheduling assumption) is similar to
choosing a weaker guard such that safety properties are satisfied: it is
minimalistic and gives more flexibility for later refinements.

\newBevt[wfevt]{\evt_{wf}}
\newBevt[schevt]{\evt_{sch}}
\newBevt[sfevt]{\evt_{sf}}
The relationship between our schedules (coarse/fine) and fairness
assumptions (weak/strong) can be illustrated as follows.  Consider
the following events with identical actions. The guard of these events
are also the same as $\csched \land \fsched$ for some predicates
$\csched$, $\fsched$.
\begin{align*}
  & \ubeventinlineidx{\wfevt}{}%
  {\csched \land \fsched}{}{}{\csched \land \fsched}{\ldots}
  \\
  &\ubeventinlineidx{\schevt}{}%
  {\csched}{\fsched}{}{\csched \land \fsched}{\ldots}
  \\
  &\ubeventinlineidx{\sfevt}{}%
  {}{\csched \land \fsched}{}{\csched \land \fsched}{\ldots}
\end{align*}
Event \wfevt is scheduled with weakly fairness, event \sfevt is
scheduled with strongly fairness.  Event \schevt's scheduling is split
between a coarse schedule \csched and a fine schedule \fsched.
Consider the strength of their scheduling assumption, we have the
following relationship:
\[\schedule.\wfevt \Wide\follows \schedule.\schevt \Wide\follows
\schedule.\sfevt~.\] In fact, using coarse and fine schedules, we can
specify a finer-grained spectrum of scheduling assumptions (compared
to fairness assumptions) with the minimum being the weak-fairness
assumption and the maximum being the strong-fairness assumption, as
can be seen in Figure~\ref{fig:schedules}.
\begin{figure}[!htbp]
  \centering
  \ifx\PREAMBLE\UnDef
\documentclass{beamer}
\usepackage{tikz}
\usetikzlibrary{shapes}
\usepackage[english]{babel}

\usepackage[latin1]{inputenc}
\usepackage[color]{eventB}
\usepackage{sgnctrl}

\begin{document}
\else
\fi
\begin{tikzpicture}[scale=1]
  \draw (0,0) node(wf)[align=center]{Weak\\Fairness};
  \draw (6, 0) node(sf)[align=center]{Strong\\Fairness};
  \draw [->] (wf) --node[above, fill=white, align=center]{Coarse/Fine\\Schedules} (sf);
\end{tikzpicture}

\ifx\PREAMBLE\UnDef
\end{document}
\else
\fi
  \caption{The spectrum of scheduling assumptions}
  \label{fig:schedules}
\end{figure}
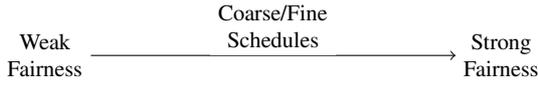
\subsection{Third Refinement --- The Actuators}
\label{sec:third-refinement}

In this refinement \MchIII, we focus on requirements \ref{eqp:sgn} and \ref{env:obey} 
which describe the light signals in the station and 
state the assumption that the train drivers always obey these light signals. So 
far, \moveout and \movein which model the behaviour of individual 
trains and their driver, state that the trains move when it is safe to. 
However, the drivers often cannot judge for themselves whether it is 
safe to proceed because they cannot see all the dangers. This is why 
we adopt the convention of the light signals: wherever it may be 
dangerous to proceed, a light signal is located and can only be green
if it is safe for the train it is addressed to advance. This allows us
to change the guard and schedules of the train events to only refer to 
information that the drivers have access to.

We continue our development by modelling the signals associated 
with different blocks within the network.  Variable \signal is introduced 
to denote the set of platform for which the light signal is green.  We 
focus the rest of this section on the control of the signals regulating 
the departure from the platforms.  In particular, invariants 
\Binv{inv3\_2} and \Binv{inv3\_3} 
state that if a platform signal is \emph{green} then the exit block is 
free and the other platform signals are \emph{red}. Invariant \Binv{inv3\_4} 
states that the signal is green only for occupied platforms.
\begin{Bcode}
  $%
  \invariants{%
    \Binv{inv3\_1}: & \signal \subseteq \PLATFORM \\
    \label{eq:inv:34}
    \Binv{inv3\_2}: & \Exit \in \ran.\location \1\implies \signal = \emptyset  \\
    \Binv{inv3\_3}: & \qforall{p,q}{p, q \in \signal}{p = q} \\
    \Binv{inv3\_4}: & \signal \1\subseteq \ran.\location } $
\end{Bcode}

We refine the \moveout event to use the platform signal as follows.
\begin{Bcode}
  $%
  \ubeventidx{\moveout}{\train}{}%
  {%
    \quad \train \in \trains \land
    \location.\train \in \PLATFORM \\ \land ~
    \setlength{\fboxsep}{2pt}
    \framebox{$\location.\train \in \signal$}
  }%
  {%
    \quad \train \in \trains \land \location.\train \in \PLATFORM \\ 
    \setlength{\fboxsep}{2pt}
    \land~ \framebox{$\location.\train \in \signal$}}
  { \msout{\neg\,\Exit \in \ran.\location} }
  {\location.\train \bcmeq \Exit \\
    \signal \bcmeq \signal \setminus \{ \location.\train \}%
  }%
  $
\end{Bcode}
The refinement of \moveout is justified by applying
Theorem~\ref{thm:schedule-ref} which requires us to prove the 
following:
\begin{gather}
\tag{C\_FLW\_3} \label{eq:c-flw3}
	\langle \location.\train \in 
		\PLATFORM \rangle
	\WIDE\leadsto \langle \location.\train \in 
		\PLATFORM \cap \signal \rangle  \\
\tag{C\_STB\_3} \label{eq:c-stb3}
	\langle \location.\train \in 
		\PLATFORM \cap \signal \rangle 
	\WIDE\un \neg \langle \location.\train \in 
		\PLATFORM \rangle \\
\tag{F\_FLW\_3} \label{eq:f-flw3}
	\langle \location.\train \in 
		\PLATFORM \rangle 
		\land \neg \, \Exit \in \ran.\location
    \WIDE\leadsto \true \\
\tag{F\_STR\_3} \label{eq:f-str3}
	\begin{array}{@{}rl}
	&\langle \location.\train \in 
		 \PLATFORM \cap \signal \rangle \land \true \\
	\limp~~ & \neg \, \Exit \in \ran.\location
	\end{array}
\end{gather}
\begin{sloppypar}
\eqref{eq:f-flw3} follows directly from the implication rule
\eqref{eq:implication};
\eqref{eq:f-str3} follows from \Binv{inv3\_2};
\eqref{eq:c-stb3} requires a number of simple proofs which will be left 
out.  To discharge \eqref{eq:c-flw3}, we apply the 
ensure-rule \eqref{eq:ensure}.
\end{sloppypar}
\begin{gather}
  \label{eq:saf3-6}
  \langle \location.\train \in 
		\PLATFORM \rangle \3\un 
	\langle \location.\train \in 
		\PLATFORM \cap \signal \rangle
  \tag{\Binv{un3\_1}} \\
  \label{eq:prg3-5}
  \tr ~ \langle \location.\train \in
  \PLATFORM \setminus \signal \rangle ~
  \tag{\Binv{prg3\_1}}
\end{gather}

Ignoring safety property \ref{eq:saf3-6}, we focus on
\ref{eq:prg3-5}. In order to have an event to falsify $\langle
\location.\train \in \PLATFORM \setminus \signal \rangle$, we need the
event to add $\location.\train$ to $\signal$, i.e., to turn to green
the signal of the platform where $\train$ is.  This is clearly a task
of the controller rather than a model of the trains. Therefore, we
introduce a controller event, \ctrlplf, indexed with the platforms.
Our first design for \ctrlplf is as follows.
\begin{Bcode}
  $%
  \ubeventidx{\ctrlplf}{\platform}{}%
  {%
  }%
  {
    \platform \in \ran.\location \land
    \platform \in \PLATFORM \setminus \signal
  }
  {%
  }
  {\signal \bcmeq \signal \bunion \{ \platform  \} }
  $
\end{Bcode}

The proof that \ctrlplf implements \ref{eq:prg3-5} is as follows.
\begin{calculation}
  \MchIII \Wide\vdash \tr ~ \langle \location.\train \in \PLATFORM
  \setminus \signal \rangle 
  \hint{\sqsubseteq}{ Transient rule ~\eqref{eq:tr-rule} with
    $\ctrlplf.(\location.\train)$}
  \MchIII \Wide\vdash
  \begin{array}[t]{@{}l}
    \ctrlplf.(\location.\train) ~\falsifies~\\
    \quad \langle \location.\train \in
    \PLATFORM \setminus \signal \rangle
  \end{array}
  \hint{=}{ Falsifies rule~\eqref{eq:falsifies} with \eqref{eq:C-EN:4}
    and \eqref{eq:NEG:4}}
  \true
\end{calculation}
where:
\begin{gather}
  \label{eq:C-EN:4} \tag{C\_EN\_4} 
  \begin{array}{rlll}
    & &\langle  \location.\train \in \PLATFORM \setminus \signal \rangle & \\
    \implies & & & \\
    & & \framebox{$ \location.\train \in \ran.\location $} & \textrm{\quad // coarse schedule} \\[0.5ex]
    & \land & \framebox{$\location.\train \in \PLATFORM \setminus
      \signal$}
    & 
  \end{array} \\[2ex]
  \label{eq:NEG:4} \tag{NEG\_4}
  \begin{array}{rlll}
    & & \langle  \location.\train \in \PLATFORM \setminus \signal \rangle & \\
    & \land & \location.\train \in \ran.\location & \\
    & \land & \location.\train \in \PLATFORM \setminus \signal & \\
    & \land & \framebox{$ \signal' \1= \signal \bunion \{ \location.\train \}$} &
    \textrm{\quad // action} \\[1ex]
    & \land & \framebox{$\location' \1= \location$} &  \\
    \implies \\
    & & \neg \, \langle \location'.\train \in \PLATFORM \setminus \signal' \rangle &
  \end{array}
\end{gather}
Notice that we choose the coarse schedule of \ctrlplf so as to simplify the proof
of \eqref{eq:C-EN:4}. 
Furthermore, we choose the assignment to $\signal$ in such a way as to
satisfy \eqref{eq:NEG:4}.

The next step is to prove that \ctrlplf satisfies the safety
properties (i.e. unless properties and invariants) of the current
refinement. In order to prove \Binv{inv3\_2} and \Binv{inv3\_3}, we
need to strengthen the guard to:
\begin{align*}
   \platform \in \PLATFORM 
  \1\land  \platform \in \ran.\location
  \1\land \neg \Exit \in \ran.\location 
  \1\land  \signal = \emptyset~.
\end{align*}
Due to feasibility condition~\eqref{eq:fis}, we need to strengthen the
schedules of \ctrlplf accordingly.  Since strengthening the current
coarse schedule will invalidate the current proof of \eqref{eq:C-EN:4}, we introduce the
following new fine schedule for \ctrlplf:
\[ \neg \Exit \in \ran.\location \1\land  \signal = \emptyset~.\]
Event \ctrlplf is finalized as:
\begin{Bcode}
  $%
  \ubeventidx{\ctrlplf}{\platform}{}%
  {%
    \setlength{\fboxsep}{0pt}
   \framebox{\minibox{$\,\,\quad ~ \platform \in \PLATFORM \1\land
    \platform \in \ran.\location$ \\
    $\,\,\land\ ~ \neg \Exit \in \ran.\location \1\land 
    \signal = \emptyset$
    }}
  }%
  {
    \platform \in \ran.\location \land
    \platform \in \PLATFORM \setminus \signal
  }
  {%
   \setlength{\fboxsep}{2pt}
   \framebox{$ \neg \Exit \in \ran.\location \land
    \signal = \emptyset$}
  }
  {\signal \bcmeq \signal \bunion \{ \platform  \} }
  $
\end{Bcode}
Table~\ref{table:ctrlplf:design-guideline} summarizes the design choices behind event \ctrlplf.
\begin{table}
\begin{tabular}{|l|l|l|}
    \hline
    Event part & Formula & PO
 \\ \hline
    \hline
    coarse schedule & \Binv{prg3\_1} & \eqref{eq:C-EN:4} 
 \\ action & \Binv{prg3\_1} & \eqref{eq:NEG:4}
 \\ guard & \Binv{inv3\_2} and \Binv{inv3\_3} & invariance 
 \\ fine schedule & guard & \eqref{eq:fis} 
 \\ \hline
\end{tabular}
\caption{Proof obligations and formulas justifying the design of \ctrlplf }
\label{table:ctrlplf:design-guideline}
\end{table}

It is interesting to see that, in this refinement, we effectively
shift the fine schedule from \moveout (an environment event) to
\ctrlplf (a controller event). The model remains abstract when it
comes to specify the order in which the trains gain access to the exit
block but specific enough to maintain liveness.

Event \ctrlplf is a specification for the computer to control the
platform signals satisfying both safety and liveness properties of the
overall system.  In particular, the scheduling information states that 
if (1) a platform is occupied and the platform signal is \emph{red} 
infinitely long and (2) the exit block is unoccupied and the other 
platform signals are all \emph{red} infinitely often, then the system 
should eventually turn this platform signal to \emph{green}. The 
refinement of event \movein and how the entry signal is controlled
is similar and omitted for the rest of the paper.

The new fine schedule changes the earlier proof of \ref{eq:prg3-5} in
two ways.  First, the fine schedule gets in the antecedent of 
\eqref{eq:NEG:4}, which does not invalidate the proof of 
\eqref{eq:NEG:4}.  Second, we get an additional leads-to property to
prove corresponding to \eqref{eq:F-EN}.
\begin{gather}
\tag{F\_EN\_4} \label{eq:F-EN:4}
\langle  \location.\train \in
\PLATFORM \setminus \signal \rangle
\WIDE\leadsto 
\begin{array}[t]{rl}
  &\neg\,\Exit \in \ran.\location \\
  \land&  \signal = \emptyset 
\end{array}
\end{gather}

We start the proof of \eqref{eq:F-EN:4} by weakening its right-hand
side to $\true$ (consequence of transitivity~\eqref{eq:transitivity}
and implication~\eqref{eq:implication} rules) and we keep refining it
until we can implement it simply.
\begin{calculation}
   	\true \wide\leadsto 
		\neg\,\Exit \in \ran.\location 
	    \1\land  \signal = \emptyset 
\hint{\sqsubseteq}{ \eqref{eq:transitivity} }
   	\true \Wide\leadsto 
	    \signal = \emptyset 
\\ &   	\signal = \emptyset  \Wide\leadsto 
		\neg\,\Exit \in \ran.\location 
	    \1\land  \signal = \emptyset 
\mhint[.68\columnwidth]{\sqsubseteq}{ \eqref{eq:psp} on second 
	with $p := \true$~, 
		$q := \neg\,\Exit \in \ran.\location$~,
		$r := \signal = \emptyset$~,
		$b := \neg\,\Exit \in \ran.\location 
	    \1\land  \signal = \emptyset$ }
          \begin{array}[t]{@{}lr}
            \true \wide\leadsto \signal = \emptyset & (\Binv{prg3\_2})\\
            \true  \wide\leadsto \neg\,\Exit \in \ran.\location &
            (\Binv{prg3\_3}) \\
            \signal = \emptyset  \Wide\un
		\neg\,\Exit \in \ran.\location 
	    \1\land  \signal = \emptyset & (\Binv{un3\_2})
          \end{array}
\end{calculation}

We leave out the detailed proof for \Binv{prg3\_2} and \Binv{un3\_2}:
\Binv{prg3\_2} is a transient property which can be implemented by
\moveout event, \Binv{un3\_2} is a trivial safety property.

Property \Binv{prg3\_3} is identical to \ref{eq:prg2-3} in the second
refinement \MchII.  However, we cannot directly reuse \ref{eq:prg2-3}
since this could lead to circular reasoning.  As explained below,
additional precautions are required to avoid the problem.

\subsubsection{Reusing Progress Properties}

Property \ref{eq:prg2-3}, which states that the exit block becomes
free infinitely often, turns out to be a key abstraction in \MchII.
In \MchIII and later refinements, it will be very important that the
exit block be available infinitely many times so that, even if a given
train misses the first opportunity to move away from its platform, it
still is certain to get a turn eventually.  As a result, we would like
to reuse \ref{eq:prg2-3} in the reasoning about \MchIII and 
subsequent refinements.

Earlier, we have proved that \MchII satisfies \ref{eq:prg2-3}, i.e.,
\begin{equation}
\ew{\execution.\MchII \wide\limp \ref{eq:prg2-3}}~.\label{eq:prg2-3:proof}
\end{equation}
To ensure that \MchIII refines \MchII, we prove
\begin{equation}
\ew{ \execution.\MchIII \wide\limp \execution.\MchII\label{eq:ref3-2} }
\end{equation}
Succeed in proving \eqref{eq:ref3-2}, together with
\eqref{eq:prg2-3:proof}, will ensure that \MchIII also
satisfies \ref{eq:prg2-3}.
However, we cannot directly reuse \ref{eq:prg2-3} during the proof of
\eqref{eq:ref3-2}: \emph{our reasoning would be circular}.

In order to avoid circular reasoning, for each model, we keep a binary
relation representing the dependency between progress properties of
interest and the events that contribute to their implementation.
Consider the initial model \InitM, since the property
\ref{eq:prg0-1} is eventually implemented by \depart, the dependency
relation can be
\[\{\ref{eq:prg0-1} \mapsto (\InitM.)\depart\}.~\]
\begin{sloppypar}
In the first refinement, consider the dependency for \ref{eq:prg0-1},
it is dependent on event $(\MchI.)\depart$ (which is the refinement of the abstract
event $(\InitM.)\depart$) and events \moveout, \movein (which are used for
proving the refinement of $(\InitM.)\depart$ by $(\MchI.)\depart)$.
\end{sloppypar}
\begin{align*}
	\{ & \ref{eq:prg0-1} \mapsto (\MchI.)\depart,  \\
	 & \ref{eq:prg0-1} \mapsto (\MchI.)\moveout, \\
	 & \ref{eq:prg0-1} \mapsto (\MchI.)\movein\}~.
\end{align*}

More formally, the dependency relation denotes the
relationship between the scheduling assumptions of the events and the
property these events implement.  Consider a model \Mch and its 
dependency relation:
\[\{P_1 \mapsto (\Mch.)\evt_1, 
	P_1 \mapsto (\Mch.)\evt_2, 
	P_2 \mapsto (\Mch.)\evt_3\}~.\]
The above relation encodes the following conditions:
\begin{align}
  \ew{ \safety.\Mch \land \schedule.\evt_1 \land \schedule.\evt_2
  & \wide\limp P_1 } \\
  \ew{ \safety.\Mch \land \schedule.\evt_3
  & \wide\limp P_2 }
\end{align}
For a given model, we only need to consider the dependency 
relation for the progress properties that we want subsequent 
refinements to reuse.  By default, progress properties are
not reused.



In summary, the dependency relation summarizes the proofs
(accumulated through refinement) of progress properties.
We use it in order to avoid any circular reasoning linked to the reuse
of progress properties.  A progress property can be reused only after
its effecting events have been refined.  Naturally, during
the proof of refinements of these dependent events, the progress
property cannot be used.  This means that if, in a refinement of
$\Mch$ in the previous example, we replace the coarse schedule of
$\evt_1$, we cannot use $P_1$ in the proof of \eqref{eq:csr-c-flw}.

Coming back to the train station example, we are interested in reusing
\ref{eq:prg2-3} which is introduced in \MchII.  Since \ref{eq:prg2-3}
is implemented by \depart, the dependency relation for \MchII is as
follows:
\[\{ \ref{eq:prg2-3} \mapsto (\MchII.)\depart \}~.\]%
Since \depart is unchanged in \MchIII, its refinement is trivial.
Therefore, we are free to make use of \ref{eq:prg2-3} in all the
proofs of liveness in \MchIII.
%
%
It also follows that, the dependency for \MchIII is
the same as that of \MchII, i.e.,
\[\{ \ref{eq:prg2-3} \mapsto (\MchIII.)\depart \}~.\]
In fact, property \ref{eq:prg2-3} is also reused in future
refinements.

\subsection{Fourth Refinement --- The Controller}
\label{sec:fourth-refinement}

In this refinement \MchIV, we focus on realising the software
controller.  At the end of the previous refinement, the controller is
entirely specified by \ctrlplf which has a fine schedule. Although the
fine schedule is a useful specification mechanism, we argue that it is
not readily implementable.  While it is easy to produce a correct (if
not efficient) implementation for an event that has only a coarse
schedule --- a program that tests infinitely often (every second,
every minutes or every year) the schedule and execute the event when
its guard is true would be a correct implementation --- it is not so
straightforward for a fine schedule. Repeatedly testing a fine
schedule will be incorrect in a situation where the fine schedule
becomes true and false infinitely many times and that the program just
happens to test infinitely many times only when it is false.  This
naive scheduler will fail to detect that the event has to be executed.

%
To make our controller more deterministic, we now proceed to refining
\ctrlplf's fine schedule away. 
For that purpose, we introduce three new variables ($\queue$, $\head$,
$\tail$) to model a queue. Variable $\queue$ is an injective function from
platform to an interval of integers where $\head$ is the index of the
first element of the queue and $\tail$ is the index just after the
last element of the queue. This entails that the queue is empty when
$\head = \tail$.

As a convention, platforms
are pulled at $\head$ and inserted at $\tail$. Below, $[\head, \tail)$
is an integer interval that includes $\head$ but excludes $\tail$ and $
(\queue^{-1})[~\{ \head \}~] $ is the application of the inverse of $
\queue$ to the set $\{ \head \}$. The latter has the particularity that, 
when $\head$ is not in the range of $\queue$, the expression 
evaluates to the empty set rather than being undefined.
\begin{Bcode}
  $ \invariants{ \Binv{inv4\_1}: & \queue \in \PLATFORM \pinj [\head, \tail)  \\
    \Binv{inv4\_2}: & { \signal \subseteq (\queue^{-1}) [~\{ \head \}~] }  \\
    \Binv{inv4\_3}: & \dom.\queue \1= \PLATFORM \cap \ran.\location } $
\end{Bcode}
In order to maintain \Binv{inv4\_3}, we let \movein increase $\tail$ and 
insert the platforms in the queue as they become occupied and 
\moveout increase $\head$ and remove the platforms as they become 
free.
   \begin{Bcode}[\small]
   $
   \ubeventidx{\movein}{\train}{}
   { ...
   }
   { ... }
   {}
   { ... \\
 	\tail \bcmeq \tail + 1 \\
 	\queue.(\location'.\train)\! \bcmeq\! \tail  \\ 
	\location.\train \bcmin \PLATFORM \\
	\qquad \quad \setminus \ran.\location \\
	}
   $
   $
   \ubeventidx{\moveout}{\train}{}
   { \location.\train \in \signal
   }
   { ... }
   {}
   { ... \\
 	\head \bcmeq \head + 1 \\
 	\queue\! \bcmeq\! \queue \ransub \{ \head \}   \\
	\location.\train \bcmeq \Exit \\
	\signal \bcmeq \signal \\
	\qquad \quad \setminus \{ \location.\train \}
   }
   $
 \end{Bcode}

With the queue, we can schedule the controller deterministically: to
turn green the light signal of a platform, that platform has to be
the head of the queue.  Event \ctrlplf is refined as follows.
\begin{Bcode}
  $
  \ubeventidx{\ctrlplf}{\platform}{}{
    \quad \platform \in \PLATFORM \land
    \platform \in \ran.\location \\ \land 
    ~ \neg \Exit \in \ran.\location \land 
    \signal = \emptyset
  }
  {
    \quad  \msout{\platform \in \PLATFORM \cap \ran.\location }  \\
    \quad \platform \in \dom.\queue \\ 
    \land ~ \queue.\platform = \head \\ 
    \land ~ \neg \Exit \in \ran.\location \\ 
    \land ~ \neg \platform \in \signal
  }
  {
    \msout{ \neg \Exit \in \ran.\location \land
      \signal = \emptyset }
  }
  { \signal \bcmeq \signal \bunion \{ \platform  \} }
  $
\end{Bcode}

\begin{sloppypar}
We apply \Thm~\ref{thm:schedule-ref} to prove the refinement of
\ctrlplf.  Omitting the trivial obligations related to
\eqref{ref:f-flw} and \eqref{ref:f-str}, we focus on the obligations
for replacing $\platform \in \PLATFORM \cap \ran.\location$ (which is
equivalent to $\platform \in \dom.\queue$) with $\langle
\queue.\platform = \head \rangle \land \neg \Exit \in \ran.\location $
for the coarse schedule of \ctrlplf (i.e., conditions \eqref{ref:c-flw}
and \eqref{ref:c-stb}).
\end{sloppypar}
\begin{gather}
  \label{eq:c-flw:5}
  p \in \dom.\queue \WIDE\leadsto 
  \begin{array}[t]{l}
    \langle \queue.\platform = \head \rangle \wide\land \\
    \neg \Exit \in \ran.\location
  \end{array}
  \tag{C\_FLW\_5} \\
  \label{eq:c-stb:5}
  \begin{array}{rl}
  & \langle \queue.\platform = \head \rangle \\
  \wide\land & \neg \Exit \in \ran.\location 
  \WIDE\un 
  \neg p \in \dom.\queue
  \end{array}
  \tag{C\_STB\_5}
\end{gather}

So far in our development, progress properties can be separated in two
different groups:
\begin{enumerate}
\item Those that are satisfied in a single step. These properties are
  proved by transforming them into transient properties (for
  example, making use of rules such as ensure-rule \eqref{eq:ensure}).
  Each transient property is implement by an individual event using a
  combination of transient rule \eqref{eq:tr-rule} and falsifies rule
  \eqref{eq:falsifies}.
\item Those that are satisfied in some pre-determined number of
  steps.  These properties are proved by breaking them down into
  several properties that can be satisfied in a single step using
  transitivity~\eqref{eq:transitivity}.
\end{enumerate}

Property \eqref{eq:c-flw:5} does not fit neither categories so far.  In
fact, the number of steps to satisfy \eqref{eq:c-flw:5} depends on the
position of the platform \platform within the queue \queue.  As a
result, in order to prove \eqref{eq:c-flw:5} we apply the induction
rule~\eqref{eq:induction}.
\begin{calculation}
  p \in \dom.\queue \WIDE\leadsto 
  \begin{array}[t]{l}
    \langle \queue.\platform = \head \rangle \wide\land \\
    \neg \Exit \in \ran.\location
  \end{array}
  \hint{\sqsubseteq}{ Induction rule~\eqref{eq:induction}}
  \begin{array}[t]{rl}
   \langle \queue.\platform \0- \head \0= M \rangle 
  \leadsto &
    \langle \queue.\platform - \head < M \rangle \wide\lor\\
    & \langle \queue.\platform \0= \head 
    \land \neg \Exit \in \ran.\location\rangle
  \end{array}
  (\Binv{prg4\_1})
  \mhint[.75\columnwidth]{\sqsubseteq}{ \eqref{eq:psp} with
                $p := \langle \queue.\platform - \head = M \rangle$~, 
                $q := \neg \langle \queue.\platform - \head = M \rangle $~,
		$r := \langle \queue.\platform - \head \le M \rangle$~,
		$b := \langle \queue.\platform = \head \rangle
                \land \neg \Exit \in \ran.\location$ }
   \begin{array}{lr}
     \langle \queue.\platform - \head = M \rangle \WIDE\leadsto 
     \neg \langle \queue.\platform - \head = M \rangle &
     (\Binv{prg4\_2}) \\[1ex]
     \begin{array}{@{}rl}
       \langle \queue.\platform - \head \le M \rangle 
       \WIDE\un &
       \langle \queue.\platform = \head \rangle 
       \1\land \\ & \neg \Exit \in \ran.\location
     \end{array} &
     \quad \quad (\Binv{un4\_1})
   \end{array}
\end{calculation}
{
%

%



We focus on the development for \Binv{prg4\_2}. It basically says that,
eventually, either the value of $\queue.\platform - \head$ changes or
$\platform$ is no longer in the queue.  While this is exactly what
\moveout does, we cannot prove that \moveout falsifies $\langle
\queue.\platform - \head = M \rangle$.  This is because it could take as
many as three steps to do that.  For example, let us assume that there
is a train at the \Exit block, i.e., $\Exit \in \ran.\location$, all the
platform signals are red, hence $\signal = \emptyset$.  In order to
falsify $\langle \queue.\platform - \head = M \rangle$, the following
steps have to happen:
\begin{enumerate}
\item Event \depart frees the \Exit block,
\item Event \ctrlplf turns the platform signal to green for some 
  platform,
\item Event \moveout moves to the exit block the train located at the 
  platform for which the signal is green.
\end{enumerate}

As a result, we use \eqref{eq:transitivity} to split
\Binv{prg4\_2} into three different properties.
\begin{gather}
  \label{eq:prg:4:6}
  \begin{array}{rl}
   \langle \queue.\platform - \head = M \rangle 
    \wide\leadsto &
    \langle \queue.\platform - \head = M \rangle \1\land \\
    & \neg \Exit \in \ran.\location
    \end{array}
  \tag{\Binv{prg4\_3}} \\
  \label{eq:prg:4:7}
  \begin{array}{rl}
  & \langle \queue.\platform - \head = M \rangle \\
    \1\land & \neg \Exit \in \ran.\location 
  \end{array}
    \wide\leadsto
   \begin{array}{rl}
    & \langle \queue.\platform - \head = M \rangle \2\land \\
    & \neg \, \signal  \subseteq \emptyset  \2\land \\
    & \neg \, \Exit \in \ran.\location
   \end{array}
  \tag{\Binv{prg4\_4}} \\
  \label{eq:prg:4:8}
    \begin{array}{rll}
      & \langle \queue.\platform - \head = M \rangle & \\
    \1\land & \neg \, \signal \subseteq \emptyset  & \\
    \1\land & \neg \, \Exit \in \ran.\location
  \wide\leadsto 
    \neg \langle \queue.\platform - \head = M \rangle
    \end{array}
  \tag{\Binv{prg4\_5}} 
\end{gather}
\begin{sloppypar}
Subsequently, our intention is to implement \ref{eq:prg:4:6} with
\depart, \ref{eq:prg:4:7} with \ctrlplf, and \ref{eq:prg:4:8} with
\moveout, according to our informal reasoning before.  The detail
proofs are left out.
\end{sloppypar}

\subsection{Refinement 5 --- Removal of the Event Indices}
\label{sec:fifth-refinement}

\begin{sloppypar}
  At the end of the fourth refinement, the controller event \ctrlplf
  is indexed with the platform \platform whose signal is going to be
  turned green.  However, since \platform is determined as the head of
  the queue \queue, we can remove the index of \ctrlplf.  The final
  version of \ctrlplf is as follows.
  \begin{Bcode} 
    $ \ubeventidx{\ctrlplf}{}{}{ ~ \neg \Exit \in \ran.\location
      \1\land \signal = \emptyset
    } { \quad \head < \tail \\ 
      \land ~ \neg \Exit \in \ran.\location \\ 
      \land ~ \neg (\queue^{-1}).\head \in \signal } {%
    } { \signal \bcmeq \signal \bunion \{ (\queue^{-1}).\head \} }
    $ \\
  \end{Bcode}
  The refinement of \ctrlplf can be justified trivially using
  \Thm~\ref{thm:idx-rmv}, with $(\queue^{-1}).\head$ as the witness
  for the removed index \platform.
\end{sloppypar}

The first-in-first-out policy may appear too rigid because it does 
not allow trains to stand still at a platform for a while when they 
are ahead of schedule. We chose to adhere to it because of its
simplicity which is a correct choice since the ability for trains to
linger is not one of the stated requirements. It would, however,
make for an interesting model, one which is outside the scope
of this paper.

%
%


\subsection{Summary}
\label{sec:summary}
Our development from \InitM to \MchV is driven by both safety and
progress concerns.  In particular, we choose on purpose to take into
account liveness requirement~\ref{live:train-leave} at \InitM.  As a
result, the need to prove and maintain progress properties justifies a
number of design decisions within our development.  We summarize the
key features and techniques of \unitb that have illustrated throughout
our case study.

\begin{description}
\item[\InitM] We introduce the basis of modelling using scheduled
  events and application of transient rule~\eqref{eq:tr-rule} to prove
  simple progress properties.

\item[\MchI] We illustrate how to refine scheduled events, and
  applications of transitivity rule~\eqref{eq:transitivity} and ensure
  rule~\eqref{eq:ensure} to prove progress properties.

\item[\MchII] We discuss the difference between coarse/fine
  schedules and weak/strong fairness.

\item[\MchIII] We illustrate how progress properties that have been
  proved in earlier abstract models can be reused through refinement.

\item[\MchIV] We compare different strategies for implementing
  progress properties: single step (ensure and transient rules),
  pre-determined number of steps (transitivity rule), arbitrary finite
  number of steps (induction rule).

\item[\MchV] We illustrate how events can be made more concrete by
  removing indices.
\end{description}
\section{Conclusion}
\label{sec:conclusion}

In this paper, we presented \unitb, a formal method inspired by
\eventB and \unity.  Our method allows systems to be developed
gradually via refinement and support reasoning about both safety and
liveness properties.  An important feature of \unitb is the notion of
coarse  and fine schedules for events.  Standard weak and strong
fairness assumptions can be expressed using these event schedules.  We
proposed and prove the soundness of refinement rules to manipulate the
coarse and fine schedules so that liveness properties are preserved
automatically (i.e., without the need to reprove them).  We
illustrated \unitb by developing a signal control system.

A key observation in \unitb is the role of event scheduling regarding
liveness properties being similar to the role of guards regarding safety
properties.  Guards prevent events from occurring in unsafe states so
that safety properties will not be violated; similarly, schedules ensure the
occurrence of events in order to satisfy liveness properties.  

Another key aspect of \unitb is the role of progress properties during
refinement: the obligation to prove new progress properties in the
application of refinement rules motivates the introduction of new
events and suggests the refinement of old events. In short, the
progress considerations guide the refinement of the system.

\paragraph{Related work}

\begin{sloppypar}
\unitb and \eventB differ mainly in the scheduling
assumptions.  In \eventB, event executions are assumed to
satisfy a \emph{minimal progress} condition: as long
as there are some enabled events, one of them will be executed non-deterministically. %
Given this assumption, certain liveness properties can be proved
for \eventB models such as \emph{progress} and
\emph{persistence}~\cite{hoang11:_reason_liven_proper_event_b}.
The
minimum progress assumption is often too weak to prove
the required set of liveness properties.
Furthermore, the liveness properties that can be proved using
minimal progress have to be reproved in later refinements to ascertain
that they still hold.
\end{sloppypar}

%
%
%
TLA+~\cite{DBLP_books_aw_Lamport2002} is another well-known formal
method based on refinement supporting reasoning about liveness
properties.  The execution of a TLA+ model is also captured as a
formula with safety and liveness sub-formulae expressed in the
Temporal Logic of Actions (TLA)~\cite{DBLP:journals/toplas/Lamport94}.
Actions in TLA+ (events in \unitb) can be scheduled with weak or
strong fairness.  Refinement in TLA+ is based on the WF2 and SF2
rules~\cite{DBLP:journals/toplas/Lamport94}. Rule WF2 allows
a weakly fair event to be refined by another weakly fair event and Rule SF2 allows
a strongly fair event to be refined by another strongly fair event.  The
refinement rule for \unitb is more general than the combination of WF2 and SF2:
during refinement, we can trade freely between the weakly fair component
(i.e., the coarse schedule) and the strongly fair component (i.e., the
fine schedule).  Moreover, liveness properties in TLA+ are considered
to be \emph{unimportant}~\cite[Chapter 8]{DBLP_books_aw_Lamport2002}.
In our opinion, developing systems satisfying liveness properties is
as important as ensuring that the systems satisfy safety
properties.  We argue that the liveness properties should
be considered from the early stages of the design. Indeed, addressing
liveness properties as an after thought in the design process will
often lead to complicated proofs, since the model is not designed with
proofs of liveness properties in mind.


See \cite{DBLP:conf/birthday/MannaP10}
for a review of the temporal logic framework developed by Manna and 
Pnueli. The authors use fair transition systems for the semantics of 
concurrent or reactive programs and temporal logic for specifying 
system properties. Rules are provided for proving response  
properties (called progress properties in this paper) 
that rely on just or compassionate transitions (equivalent to the weak and strong fairness scheduling policies) of the system for their 
validity. Although the Manna-Pnueli framework does not have  a 
progress preserving refinement calculus as in \unitb, it does have 
rules for data abstraction and compositional reasoning. 

The idea of combining different formal methods to reason about
liveness properties is also explored by other
researchers. In~\cite{DBLP:conf/ifm/MeryP13}, the authors combine
\eventB and TLA+ for proving liveness properties in population
protocols.  While refinement has been used in their development,
liveness properties are not preserved: progress properties have to be
reproved at each level of refinement.


\paragraph{Future work}
Currently, we only consider superposition refinement in \unitb where
variables are retained during refinement.  More generally, variables
can be removed and replaced by other variables during refinement (data
refinement). We are working on extending \unitb to provide rules for data
refinement.

Another important technique for coping with the difficulties in
developing complex systems is composition / decomposition and is already
a part of methods such as \eventB and \unity.  We intend to investigate
on how this technique can be added to \unitb, in particular, the role
of event scheduling during composition / decomposition.

Tool support is currently under construction under the name 
\emph{Literate \unitb}. The goal is to integrate seamlessly the activities 
of modelling, proving and documenting. We do so
by making equational proofs first class citizens in models, by taking 
\LaTeX{} source files at the input of the tool and allowing arbitrary 
interleaving of model and proof elements.
We use the Z3 SMT solver \cite{DBLP:conf/tacas/MouraB08} to 
discharge the proofs obligations and to
validate the proof steps.

The goal is to allow the user to formulate formal proofs in a clear 
manner and integrate them in the documentation of the models, letting the 
tool verify that every step of reasoning is sound or suggest where a 
lemma would be needed to justify a step. Such a tool is needed for the \unitb 
method to be practical. This tool substantially reduces the burden of
validity checking therefore allowing developers to focus on the 
software design.

As is the case with Rodin~\cite{abrial10:_rodin}, the tool for \eventB, a large percentage of 
proof obligations can be discharged automatically, freeing the user 
from the need to check many simple facts. This leaves him with the job 
of proving only the hardest obligations. It is useful then to be able to 
design and present the proof of these hard theorems using a format 
that is both readable by humans and amenable to formal reasoning by 
humans. We believe the equational format 
\cite{DBLP:books/sp/GriesS93} exhibits these properties since it 
permits the user to focus on one line of reasoning at a time.



\section*{Acknowledgment}

We would like to thank the anonymous reviewers for their constructive
comments which helped improve the paper significantly. The first and
last authors gratefully acknowledge a Discovery Grant from NSERC
(National Science and Engineering Research Council). %

\bibliographystyle{spmpsci}
\bibliography{progress}

\newpage

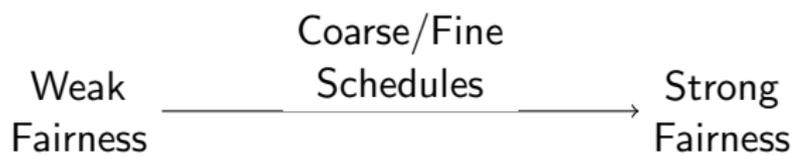
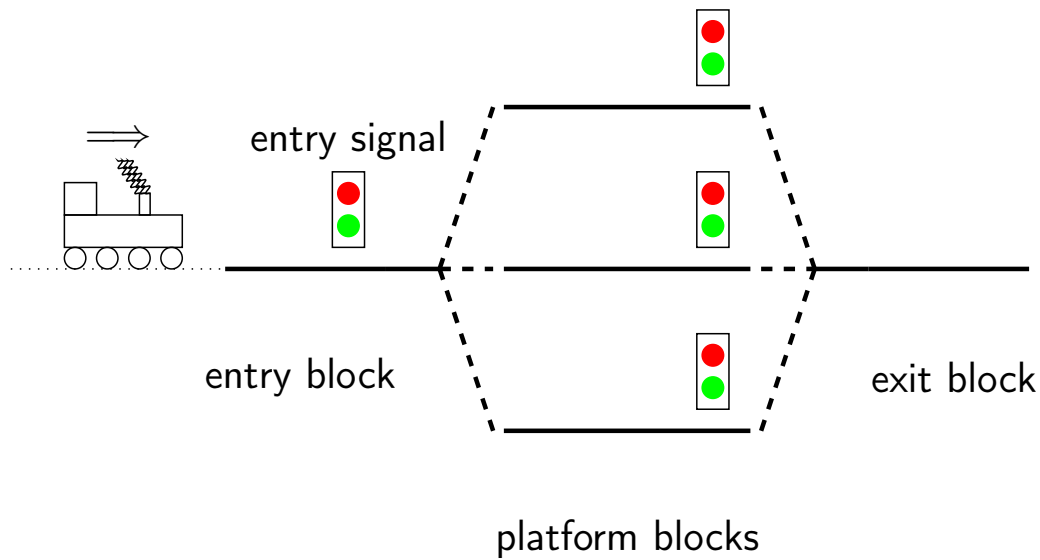
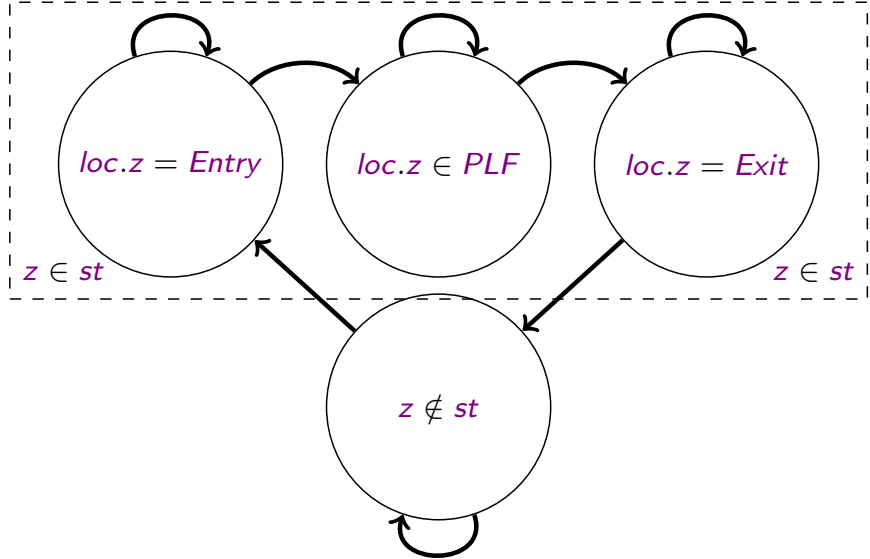
\end{document}